\documentclass[10pt,conference]{IEEEtran}

\usepackage{cite}
\usepackage[T1]{fontenc}
\DeclareUnicodeCharacter{03BA}{\ensuremath{\kappa}}
\usepackage{amsmath,amssymb,amsfonts}
\usepackage{graphicx}
\usepackage[hyphens]{url}
\usepackage{hyperref}
\usepackage{multirow}
\usepackage{booktabs}

\hypersetup{
  hidelinks,
  pdftitle={Fovea: Physical-Implication-Aware Wafer-Scale DSE with Decision-Domain-Guided Cross-Fidelity Refinement},
  pdfauthor={Jinxi Li, Huizheng Wang, Jinyi Deng, Yang Hu, and Shouyi Yin}
}

\newcommand{\workname}{Fovea}
\newcommand{\curvename}{DDC}

\title{{\workname}: Physical-Implication-Aware Wafer-Scale DSE with Decision-Domain-Guided Cross-Fidelity Refinement}

\author{
  \IEEEauthorblockN{Jinxi Li, Huizheng Wang, Jinyi Deng, Yang Hu, and Shouyi Yin}
  \IEEEauthorblockA{
    Tsinghua University, Beijing, China\\
    \texttt{ljx23@mails.tsinghua.edu.cn, wanghz22@mails.tsinghua.edu.cn,}\\
    \texttt{dengjinyi@mail.tsinghua.edu.cn, hu\_yang@tsinghua.edu.cn, yinsy@tsinghua.edu.cn}
  }
}

\begin{document}
\maketitle

\begin{abstract}

Modern pre-silicon design-space exploration (DSE) follows a coarse-to-fine workflow: low-cost evaluators screen candidate spaces, while detailed evaluation is reserved for a shortlist. Wafer-scale systems strain both stages. Architecture choices induce coupled physical implications for reticle compliance, wafer tiling, die area, D2D capability, boundary access, and placement, so the design space cannot be treated as an unconstrained Cartesian product. Meanwhile, detailed evaluation is too expensive to cover the resulting space, whereas analytical-to-reference ranking inversions make a fixed shortlist unreliable.

We present {\workname}, a reusable methodology for workload-specific wafer architecture selection rather than a fixed wafer template. 
{\workname} first performs \emph{Physical-implication-aware design-space formulation} to construct a distinct modeled-feasible space while preserving cross-dimensional trade-offs and applying only evaluator-preserving local reductions.
It then performs \emph{Decision-Domain-guided cross-fidelity refinement}. 
Paired in-domain calibration estimates workload- and space-specific analytical-to-reference disagreement, which parameterizes reference-consistent performance intervals and induces a Decision Domain for selective designated-reference evaluation. Under a valid domain-wide disagreement bound, this domain contains the designated-reference optimum; the sampling-based implementation is evaluated empirically on exhaustive-reference design spaces.
Across ten reference-verifiable design spaces and seven LLM-training workloads, {\workname} with 10\% paired calibration recovers the exhaustive designated-reference optimum in all 70 evaluated pairs while achieving $4.13\times$ average and $7.80\times$ maximum end-to-end speedup.

\end{abstract}

\section{Introduction}

The continued scaling of AI models is increasing the aggregate compute, memory, and communication demands of modern AI systems. Wafer-scale computing has consequently advanced from an architectural concept toward practical AI infrastructure, including its recent integration into OpenAI's serving stack for low-latency inference~\cite{OpenAI_Cerebras_2026}. Wafer-scale chips (WSCs) address these demands by integrating resources across a wafer-sized substrate. This extends tightly coupled integration beyond the reticle limit of a monolithic die, providing substantially greater aggregate compute and memory resources together with high-bandwidth on-wafer communication.

This growing practical relevance makes wafer-scale architecture decisions increasingly consequential, but does not imply a settled architecture. Wafer scale defines an integration canvas rather than a canonical architecture. Even within a homogeneous repeated-die organization, candidates differ in die dimensions, compatible row--column tilings, per-die compute and memory provisioning, die-to-die (D2D) capability, and functional-area allocation. Different workloads favor different compute, memory, and communication balances. Wafer-scale DSE is therefore a recurring process for selecting an architecture instance, rather than a one-time search for a universal wafer template.

Modern pre-silicon architecture exploration typically follows a multi-stage workflow: architects define a candidate space, low-cost models screen it, and detailed simulation or implementation feedback evaluates a shortlist. This workflow assumes that architectural choices map to a manageable, implementation-meaningful space and that low-cost screening retains the design preferred by detailed evaluation. Wafer-scale DSE strains both assumptions: the feasible space must first be constructed from coupled physical implications, and analytical ranking alone cannot safely prune it.

First, wafer-scale architecture choices carry coupled physical implications. Die dimensions determine reticle compliance and compatible wafer tilings, which in turn determine die count and network geometry; they also affect realizable D2D capability. Compute, memory, interconnect, and D2D-I/O resources compete for die area, while boundary-facing interfaces additionally consume die-edge resources. Consequently, the design space cannot be treated as an unconstrained Cartesian product of resource quantities. Architecture choices must be translated into their geometry, area, capability, interface, and placement implications before performance evaluation.

Second, detailed evaluation is too expensive to cover the resulting space, whereas low-cost evaluation cannot safely define a fixed shortlist. In our setup, ASTRA-sim's analytical backend requires 2.49 seconds per design on average, while the ASTRA-sim+ns-3 designated reference backend requires 2.78 hours---a cost difference of approximately $4{,}000\times$. Although analytical evaluation preserves useful global trends, 20.96\% of candidate pairs are ranked in opposite orders by the two backends, and the reference-optimal design appears, on average, at the 10.77th percentile of the analytical ranking. A fixed analytical top-$k$ cutoff can therefore exclude the reference optimum, while exhaustive reference evaluation is prohibitively expensive.

\begin{figure}[t]
    \setlength{\abovecaptionskip}{0pt}
    \centering
    \includegraphics[width=\linewidth]{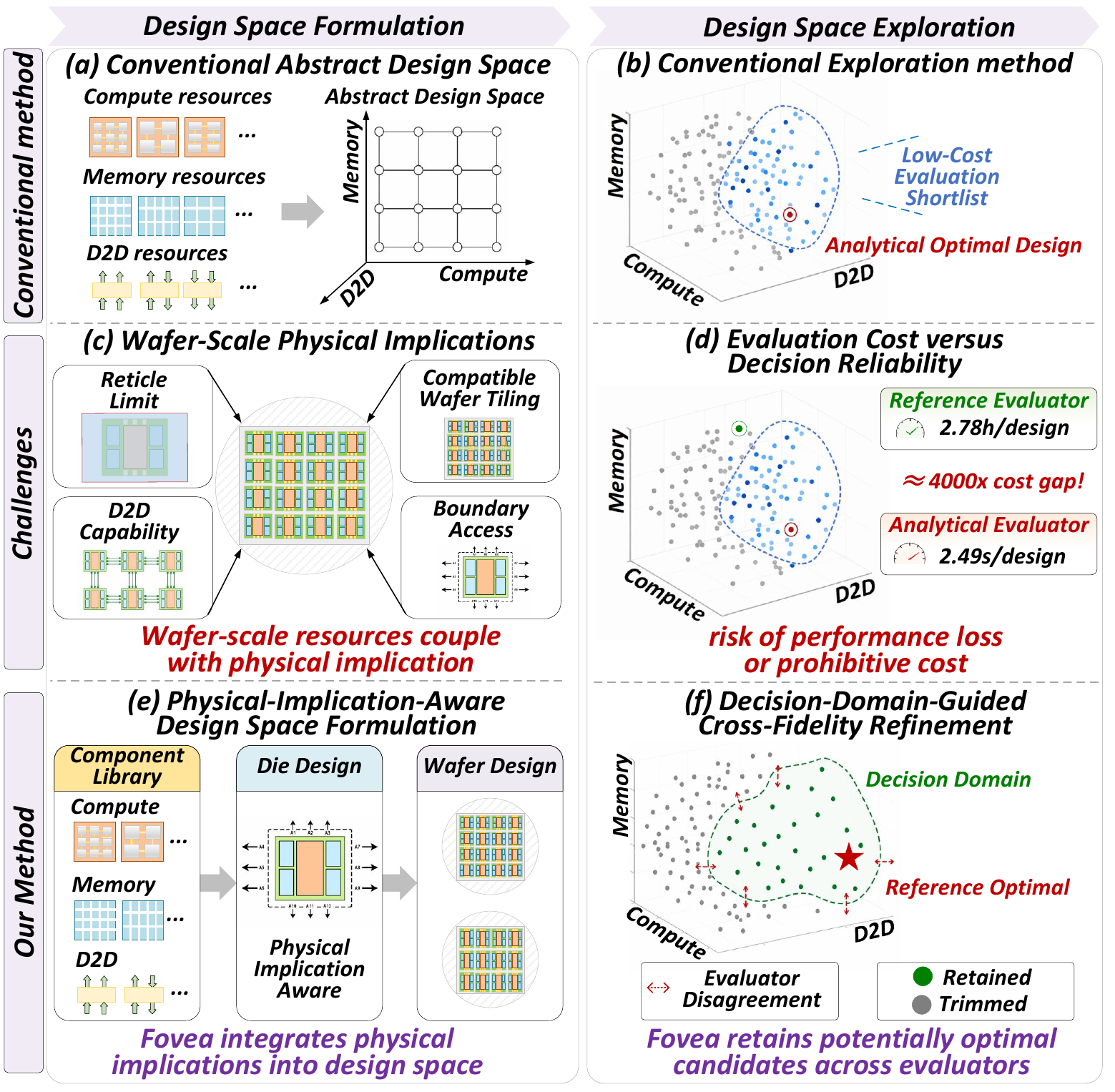}
    \caption{Wafer-scale DSE strains two assumptions of conventional multi-stage exploration. Architecture choices must first be translated into a physical-implication-aware design space, while analytical-to-reference disagreement makes a fixed low-cost shortlist unreliable. {\workname} addresses these two interfaces through physical-implication-aware design-space formulation and Decision-Domain-guided cross-fidelity refinement.}
    \label{fig:1-1}
    \vspace{-12pt}
\end{figure}

Existing search methods, including simulated annealing, evolutionary search, Bayesian optimization, and multi-fidelity optimization, can reduce evaluation cost by focusing expensive evaluations on promising candidates~\cite{Theseus,TMAC,Monad}. However, these methods primarily determine where to evaluate within a parameterized space; the search policy alone does not jointly construct a physically meaningful wafer-scale design space and derive, from workload- and space-specific evaluator disagreement, the candidate domain that still requires higher-fidelity evaluation.

We present {\workname}, a general methodology for workload-specific wafer-scale DSE. Rather than prescribing a universal wafer architecture, {\workname} provides a reusable exploration procedure that can be instantiated with a workload and optimization objective, architectural component libraries, implementation constraints, and a low-cost/reference evaluator pair. It separates wafer-scale exploration into two stages.

First, \emph{physical-implication-aware design-space formulation} maps die, compute, memory, interconnect, and D2D choices to their modeled implementation implications, constructs the distinct modeled-feasible design space under these implications, and applies exact local reductions to consolidate only evaluator-equivalent or same-footprint locally dominated configurations.

Second, \emph{Decision-Domain-guided cross-fidelity refinement} estimates analytical-to-reference disagreement through paired in-domain calibration and converts analytical scores into reference-consistent performance intervals. These intervals define a Decision Domain of candidates that cannot yet be excluded as reference-optimal, thereby guiding selective designated-reference evaluation. 
Under a valid domain-wide disagreement bound, this domain contains the designated-reference optimum. The sampling-based implementation is evaluated separately through held-out design-space validation.

Across ten reference-verifiable design spaces and seven LLM-training workloads, {\workname} with 10\% paired calibration recovers the exhaustive designated-reference optimum in all 70 workload/design-space pairs while achieving $4.13\times$ average and $7.80\times$ maximum end-to-end speedup. Its physical-implication-aware formulation excludes 29.4\% of analytically top-10\% area-feasible candidates on average and forms a distinct modeled-feasible space containing 46,782 candidates in a broader construction sweep.

This paper makes the following contributions:
\begin{itemize}
    \item \textbf{General Methodology for Workload-Specific Wafer-Scale DSE.} {\workname} decomposes wafer-scale exploration into modeled-feasible design-space construction and workload- and space-specific Decision-Domain refinement, instantiated from user-supplied workloads, components, constraints, and evaluators.

    \item \textbf{Physical-Implication-Aware Design-Space Formulation.} 
    {\workname} maps architecture choices to coupled physical implications and constructs distinct modeled-feasible spaces while preserving cross-dimensional trade-offs and removing only exact local redundancy. 

    \item \textbf{Decision-Domain-Guided Cross-Fidelity Refinement.} To the best of our knowledge, {\workname} is the first wafer-scale DSE methodology to derive an explicit error-bounded Decision Domain with conditional reference-optimum containment and use it to guide selective reference evaluation.

    \item \textbf{Reference-Verifiable End-to-End Evaluation.} Across 70 workload/design-space pairs, {\workname} recovers the exhaustive designated-reference optimum with $4.13\times$ average and $7.80\times$ maximum speedup.
\end{itemize}

\begin{figure*}[t]
    \setlength{\abovecaptionskip}{0pt}
    \centering
    \includegraphics[width=\textwidth]{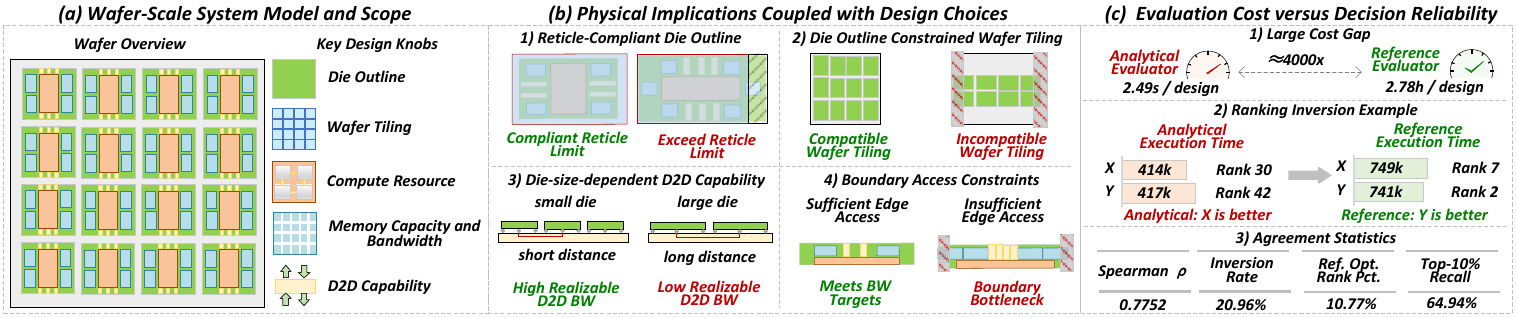}
    \caption{Wafer-scale system model, coupled physical implications, and the tension between evaluation cost and decision reliability (a) {\workname} models a homogeneous two-dimensional array of identical dies instantiated from a common reticle-compliant template. The key design choices include die outline, wafer tiling, compute resources, memory capacity and bandwidth, and D2D capability. (b) These choices are coupled through reticle, tiling, die-size-dependent D2D, and boundary-access constraints. (c) The low-cost analytical and designated ASTRA-sim+ns-3 reference backends exhibit an approximately $4{,}000\times$ cost gap, a measured ranking inversion, and imperfect decision agreement.}
    \label{fig:2-1}
    \vspace{-15pt}
\end{figure*}

\section{Wafer-Scale DSE: Background and Challenges}
\label{sec:background}

\subsection{Wafer-Scale Computing and Architectural Diversity}

Wafer-scale systems integrate compute, memory, and communication resources across a wafer-sized substrate interconnected by a high-bandwidth on-wafer fabric~\cite{Wafer-Scale,Cerebras,Dojo,SoWX2025,FRED2025}. By extending tightly coupled integration beyond the reticle limit of a monolithic die, they provide substantially greater aggregate compute throughput, memory capacity, and communication bandwidth, making wafer-scale integration a promising substrate for increasingly large AI workloads.

Wafer scale, however, defines an integration canvas rather than a canonical architecture. In the design class studied by {\workname}, the wafer is organized as a homogeneous two-dimensional array of identical, reticle-compliant dies instantiated from a common template. Each die contains compute engines, on-die memory, communication logic, and boundary-facing die-to-die (D2D) interfaces. As summarized in Fig.~\ref{fig:2-1}(a), candidate architectures differ in die dimensions, compatible row--column tilings, functional-area allocation, discrete component plans, and realizable D2D capability. The repeated-die organization fixes the overall integration structure while leaving these architectural choices open. Heterogeneous die mixtures and irregular die organizations are outside the current scope.

The preferred architecture depends on the workload and optimization objective. Compute-intensive workloads benefit from greater compute throughput, memory-sensitive workloads require different capacity and bandwidth provisioning, and communication-intensive workloads favor stronger D2D support and different array organizations. These choices compete for shared physical resources: increasing compute, memory, or D2D provisioning consumes die area, while interface resources additionally compete for boundary access; changing the die dimensions also alters the die count, network geometry, and communication distance. Section~\ref{sec:eval} shows that, under the same component libraries and physical constraints, different workloads favor different resource profiles and die organizations, and no single candidate is reference-optimal across all evaluated workloads. Training versus inference, integrated versus disaggregated deployment, and throughput versus latency optimization can introduce further design choices. Wafer-scale DSE is therefore a recurring process for selecting an architecture instance for a particular workload, objective, and technology setting, rather than a one-time search for a universal architecture.


\vspace{-5pt}
\subsection{Conventional Multi-Stage DSE Workflow}

Modern chip development commonly relies on design-space exploration, although industrial flows may refer to related activities as architecture exploration, pathfinding, trade-off analysis, design--technology co-optimization (DTCO), or system--technology co-optimization (STCO)~\cite{IBMFullSystemSimulation2006,GoogleApollo2021,TSMCDTCO2022,RochaSTCO2024}. Despite differences in terminology and implementation, these flows generally follow a multi-stage workflow: architects define the workloads, optimization objectives, design variables, and implementation constraints; low-cost analytical or trace-driven evaluators screen a broad candidate space; and detailed simulation or implementation feedback is reserved for a much smaller subset of candidates~\cite{ASTRA-SIM,Theseus,TMAC,WSC-LLM}.

The practicality of this workflow depends on two assumptions. First, architectural choices can be translated into a manageable and implementation-meaningful candidate space before expensive evaluation begins. Second, low-cost evaluation can reduce this space without excluding designs that remain competitive under detailed evaluation. Wafer-scale systems challenge both assumptions. Their architectural choices have coupled implications for die dimensions, wafer tiling, functional-area allocation, D2D capability, and die-edge resources, while analytical and reference evaluators may rank otherwise competitive candidates differently.

{\workname} addresses these two challenges through physical-implication-aware design-space formulation and Decision-Domain-guided cross-fidelity refinement.

\vspace{-5pt}
\subsection{Physical Implications of Wafer-Scale Choices}

Wafer-scale architectural choices cannot be treated as independent resource-count knobs. Die geometry, die-array organization, functional-area allocation, and D2D-I/O provisioning are coupled through geometric, area, capability, and boundary constraints. Figure~\ref{fig:2-1}(b) summarizes the physical implications captured by {\workname}'s design-space model.

\noindent\textbf{Reticle-compliant die outline.}
Lithographic exposure constrains the dimensions of an individual die~\cite{Reticle-limit,James2020PhysicalMapping}. A die therefore cannot be enlarged arbitrarily to accommodate additional architectural resources; every explored die outline must remain within the reticle envelope.

\noindent\textbf{Compatible die-array tiling.}
Die dimensions determine both the per-die resource budget and the row--column tilings that fit within the usable wafer boundary. Each compatible tiling determines the resulting die count, array geometry, and wafer coverage. Smaller dies can generally support more tiling configurations and typically yield a larger die count.

\noindent\textbf{Die-size-dependent D2D capability.}
We distinguish the fixed short inter-die channel between adjacent boundary PHYs from the on-die access wires that connect the die-level network endpoint to those PHYs. With a centered endpoint, larger outlines increase the center-to-edge wiring length. Under a fixed global-routing budget, each lane consumes more total wire and repeater resources, reducing the number of parallel lanes and hence the realizable aggregate D2D bandwidth. 

\noindent\textbf{Area budget and boundary access.}
Fixed power-delivery, clocking, control, and other implementation overheads reduce the die area available to configurable compute, on-die memory, interconnect, and D2D-I/O resources~\cite{Radhakrishnan2021PowerDelivery}. Satisfying the aggregate area budget is necessary but not sufficient for modeled feasibility: boundary-facing memory-interface and D2D-I/O blocks also require access to the die edge and may compete for the same edge-adjacent capacity~\cite{Dojo}. As Fig.~\ref{fig:2-1}(b) illustrates, a configuration may satisfy the aggregate area constraint yet fail to provide sufficient boundary access. These interface blocks must therefore satisfy both aggregate area and die-edge placement constraints.

Together, these coupled implications require an explicit modeled-feasibility formulation rather than an unconstrained Cartesian product of architectural resource quantities.

\vspace{-5pt}
\subsection{Evaluation Cost versus Decision Reliability}

We use \emph{fidelity} to denote an evaluator's modeling detail; in our evaluator pair, higher fidelity also incurs substantially greater evaluation cost. We use \emph{decision reliability} to denote agreement with the design selected by a designated reference evaluator. After the distinct modeled-feasible space has been constructed, evaluation faces a fundamental tension: a low-cost evaluator can cover the complete space but may rank candidates differently from the designated reference, whereas detailed reference evaluation provides the target design decision but is too expensive to apply exhaustively. In our implementation, ASTRA-sim's analytical backend serves as the low-cost evaluator, while ASTRA-sim with ns-3 serves as the designated reference backend~\cite{ASTRA-sim2.0,ns-3}. The latter provides the decision reference used in this study rather than silicon ground truth.

For larger workloads such as Llama-405B, the analytical backend requires 2.49 seconds per design on average, whereas the designated reference backend requires 2.78 hours, yielding an approximately $4{,}000\times$ cost difference, as summarized in Fig.~\ref{fig:2-1}(c). At these rates, evaluating the 192-design space used in the ranking-inversion example takes approximately 7.97 minutes with the analytical backend but 22.28 serial days with the reference backend. Exhaustive reference evaluation is therefore prohibitively expensive even for design spaces containing only hundreds of candidates.

This cost advantage comes with weaker decision consistency. Figure~\ref{fig:2-1}(c) shows a measured ranking inversion in the same 192-design space under an AllReduce workload with a sequence length of 2K; execution time is reported in simulator time units, and lower is better. The analytical backend predicts Design X to be faster than Design Y, with their predicted execution times differing by less than 1\%. The designated reference backend reverses this ordering and assigns Design Y a substantially better rank. Across the 70 reference-verifiable workload/design-space pairs, the two backends achieve a mean Spearman rank correlation of 0.7752, indicating that the analytical evaluator preserves useful global ordering information. Nevertheless, their mean pairwise inversion rate is 20.96\%, and the designated-reference-optimal design appears at the 10.77th percentile of the analytical ranking on average, where a lower percentile denotes a better rank. Analytical/reference top-$k$ recall, defined as the fraction of reference top-$k$ designs also present in the analytical top-$k$ set, is 57.14\%, 64.94\%, and 76.62\% for $k=5\%$, $10\%$, and $20\%$, respectively. Thus, useful global correlation does not make a fixed analytical top-$k$ cutoff consistently reliable: the designated-reference optimum can still fall outside the retained set.

Thus, analytical-only selection is scalable but may miss the designated-reference optimum, whereas exhaustive reference evaluation preserves the target decision at prohibitive cost. Fovea therefore requires a cross-fidelity exclusion rule that ties pruning to measured evaluator disagreement and reference-evaluates only unresolved candidates.

\vspace{-5pt}
\section{{\workname} Overview}
\label{sec:framework_overview}

\begin{figure}[t]
    \setlength{\abovecaptionskip}{0pt}
    \centering
    \includegraphics[width=\linewidth]{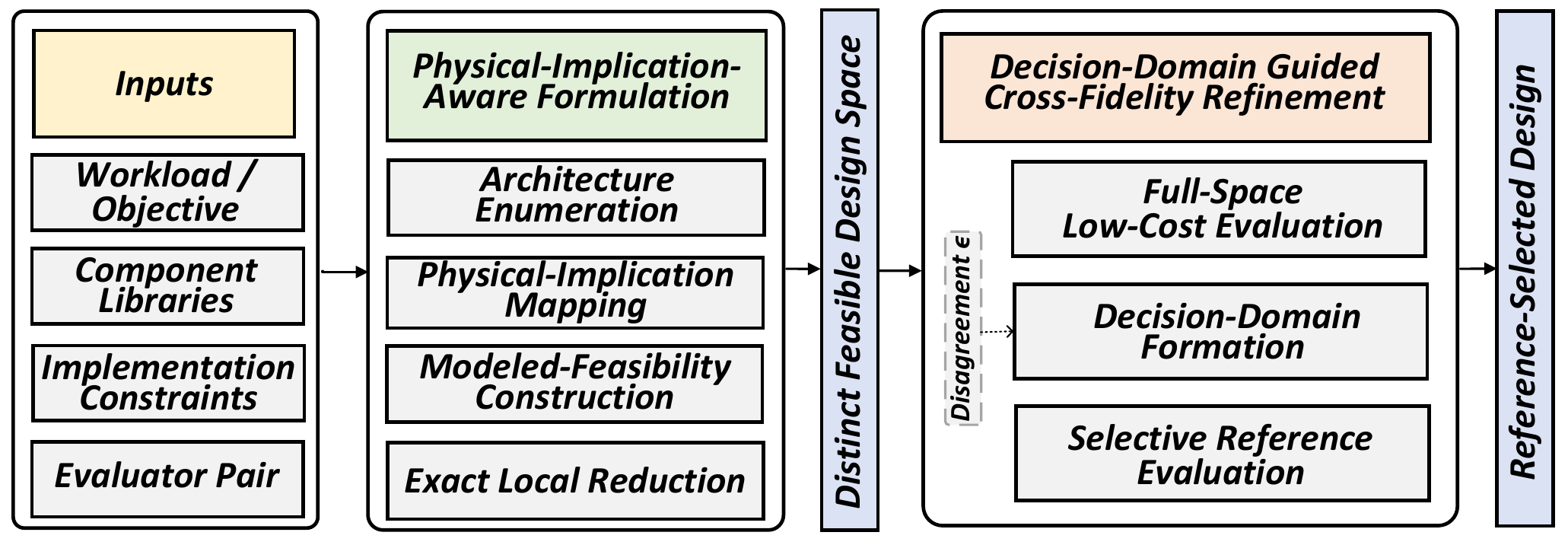}
    \caption{Overview of {\workname}. Physical-implication-aware formulation constructs a distinct modeled-feasible design space. Decision-Domain-guided cross-fidelity refinement then uses workload- and space-specific evaluator disagreement to focus designated-reference evaluation on candidates that remain potentially optimal.}
    \label{fig:4-1}
    \vspace{-18pt}
\end{figure}

As shown in Fig.~\ref{fig:4-1}, each {\workname} exploration instance is defined by four inputs: a workload and optimization objective, architectural component libraries, implementation constraints, and an evaluator pair. The first three specify the architectural choices to explore and the modeled constraints used to admit candidate combinations. The evaluator pair provides a low-cost backend for broad evaluation and a designated reference backend for resolving the retained candidates, decoupling the exploration methodology from any fixed architecture or simulator implementation.

Physical-implication-aware formulation translates the supplied architectural choices into their modeled implementation implications and retains one representative for each distinct candidate satisfying the modeled constraints. Its output is the distinct modeled-feasible wafer-scale design space $\mathcal{D}$, which defines the candidate universe for subsequent exploration.

Decision-Domain-guided cross-fidelity refinement evaluates $\mathcal{D}$ with the low-cost backend and uses workload- and space-specific evaluator disagreement to form the Decision Domain of candidates that may remain designated-reference-optimal. The designated reference backend resolves the retained candidates, and the best reference-evaluated candidate is returned as the Reference-Selected Design. Sections~\ref{sec:space_formulation} and~\ref{sec:hybrid_dse} present the two stages in detail.

\begin{figure*}[t]
    \setlength{\abovecaptionskip}{0pt}
    \centering
    \includegraphics[width=1\textwidth]{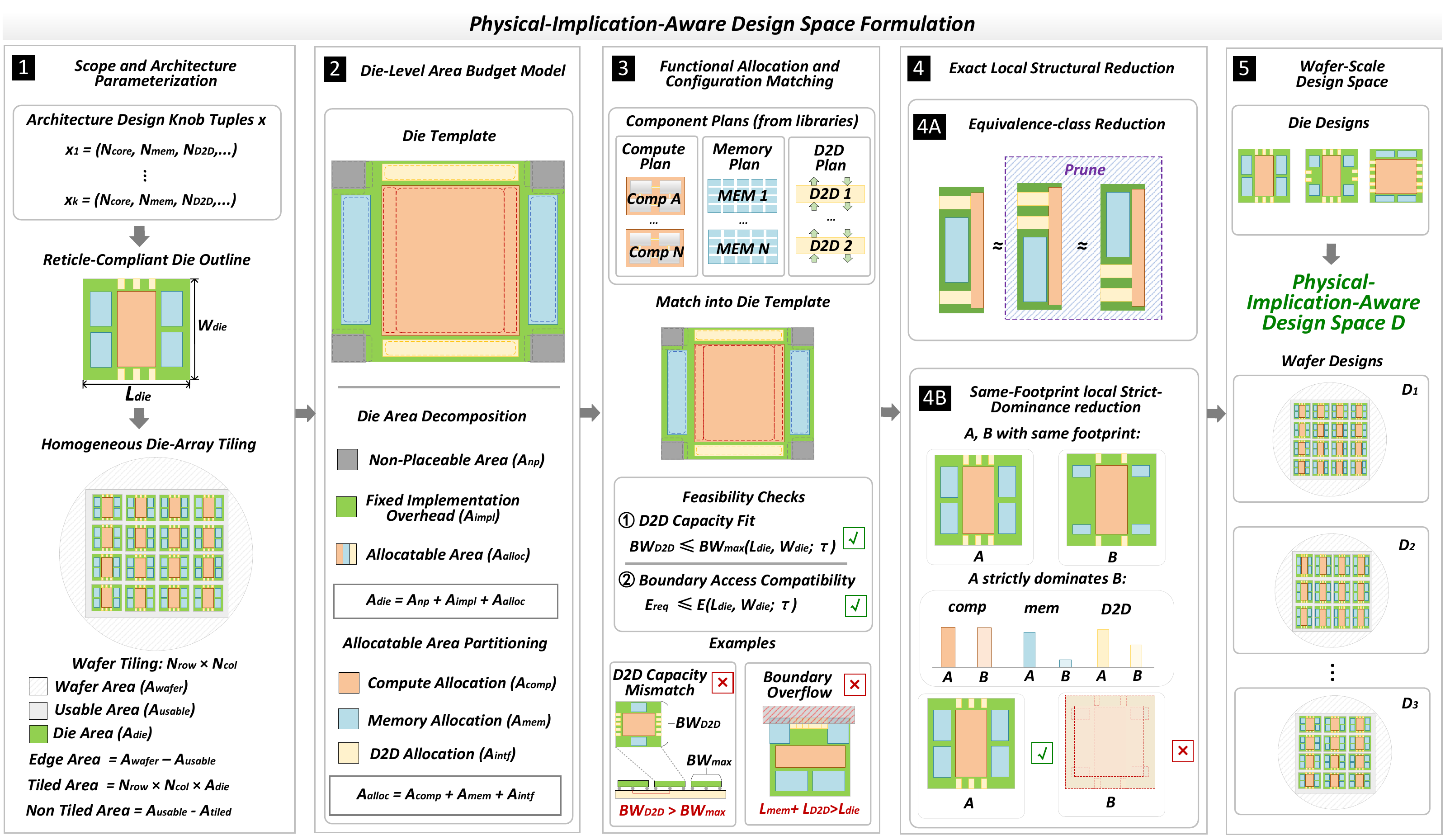}
    \caption{Physical-implication-aware design-space formulation. {\workname} enumerates reticle-compliant die outlines and compatible wafer tilings, reserves non-placeable and fixed-overhead area, and matches discrete compute, memory, interconnect, and D2D-I/O plans under coupled area, D2D-capability, boundary-access, and placement constraints. Equivalent and same-footprint locally dominated die configurations are consolidated before the retained configurations are lifted to all compatible wafer tilings, producing the distinct modeled-feasible design space $\mathcal{D}$.}
    \label{fig:5-1}
    \vspace{-17pt}
\end{figure*}

\vspace{-5pt}
\section{Physical-Implication-Aware Design-Space Formulation}
\label{sec:space_formulation}

\vspace{-5pt}
\subsection{Scope and Architecture Parameterization}

{\workname} formulates the design space within the homogeneous repeated-die architecture introduced in Section~\ref{sec:background}. Each wafer-scale candidate consists of a regular $N_{\mathrm{row}}\times N_{\mathrm{col}}$ array of identical, reticle-compliant dies connected by boundary-facing D2D interfaces. A candidate is parameterized by its die dimensions, compatible row--column tiling, functional-area allocation, and discrete compute, on-die memory, interconnect, and D2D-I/O configurations selected from the supplied component libraries.

These parameters are coupled through their implementation implications. Die dimensions determine the per-die area budget, compatible wafer tilings, adjacent-die communication distance, and realizable D2D capability, while the selected component configurations determine area demand, boundary-access requirements, and placement class. The corresponding implication functions are technology-specific inputs to {\workname}. The framework applies these functions and checks the modeled constraints before admitting an architecture-parameter tuple to $\mathcal{D}$, rather than treating the parameters as independent resource-count knobs.

\subsection{Wafer-Level Modeled-Feasibility Checks}

For each die outline $(L_{\mathrm{die}},W_{\mathrm{die}})$, Fovea first verifies reticle compliance and determines the wafer region available for regular die placement. Let $A_{\mathrm{wafer}}$ denote the total wafer area and $A_{\mathrm{edge}}$ the wafer-edge exclusion area. The usable area is
\begin{equation}
A_{\mathrm{usable}}=A_{\mathrm{wafer}}-A_{\mathrm{edge}}.
\end{equation}

Geometric compatibility is not inferred from aggregate area alone. For each integer row--column pair $(N_{\mathrm{row}},N_{\mathrm{col}})$, Fovea instantiates the corresponding regular die array after applying the declared wafer-edge exclusion and checks every die footprint against the usable-wafer boundary. A tiling is admitted only when every die lies entirely within this boundary. Thus, two arrays with the same tiled area can differ in feasibility because their row--column aspect ratios interact differently with the wafer boundary. The area quantities below are accounting attributes computed only after this geometric-containment check.

For a die area $A_{\mathrm{die}}=L_{\mathrm{die}}W_{\mathrm{die}}$, the tiled area is
\begin{equation}
A_{\mathrm{tiled}}=N_{\mathrm{row}}N_{\mathrm{col}}A_{\mathrm{die}},
\end{equation}
and the residual usable area is
\begin{equation}
A_{\mathrm{non\text{-}tiled}}=A_{\mathrm{usable}}-A_{\mathrm{tiled}}.
\end{equation}
Here, $A_{\mathrm{edge}}$ is excluded before tiling, whereas $A_{\mathrm{non\text{-}tiled}}$ records the usable area left uncovered by a particular compatible tiling. Fovea retains every tiling that passes the geometric-containment check for subsequent die-level configuration.

\vspace{-5pt}
\subsection{Die-Level Area-Budget Model}

For each die geometry, the total die area $A_{\mathrm{die}}=L_{\mathrm{die}}W_{\mathrm{die}}$ is decomposed into three aggregate area budgets: non-placeable area $A_{\mathrm{np}}$, fixed implementation overhead $A_{\mathrm{impl}}$, and allocatable area $A_{\mathrm{alloc}}$. $A_{\mathrm{np}}$ accounts for geometric loss, keep-out regions, and placement blockages; $A_{\mathrm{impl}}$ accounts for power delivery, clock distribution, control, and other fixed implementation support; and $A_{\mathrm{alloc}}$ is available for configurable architectural resources. The die-area budget satisfies
\begin{equation}
A_{\mathrm{die}}
=
A_{\mathrm{np}}
+
A_{\mathrm{impl}}
+
A_{\mathrm{alloc}}.
\end{equation}
{\workname} partitions $A_{\mathrm{alloc}}$ among compute, on-die memory, and interconnect and D2D-I/O resources:
\begin{equation}
A_{\mathrm{alloc}}
=
A_{\mathrm{comp}}
+
A_{\mathrm{mem}}
+
A_{\mathrm{intf}},
\end{equation}
where $A_{\mathrm{comp}}$, $A_{\mathrm{mem}}$, and $A_{\mathrm{intf}}$ denote the corresponding functional-area allocations. These quantities represent aggregate area consumption rather than necessarily contiguous physical regions. Component tuples satisfying the area budgets proceed to the capability and boundary-access checks described next.

\vspace{-5pt}
\subsection{Functional Allocation and Configuration Matching}

For each functional-area allocation, Fovea selects discrete compute, on-die memory, interconnect, and D2D-I/O configurations from the supplied component libraries and checks their modeled physical compatibility.

\noindent\textbf{Die-size-dependent D2D-capability check.}
For each selected D2D-I/O plan, Fovea compares its requested aggregate bandwidth against the capability envelope supported by the candidate die outline:
\begin{equation}
BW_{\mathrm{D2D}}
\le
BW_{\max}
\left(L_{\mathrm{die}},W_{\mathrm{die}};\tau\right),
\end{equation}
where $\tau$ denotes the D2D setting: per-wire rate, slice footprint, endpoint location, unit-length access cost, and routing/interface budgets. The inter-die channel between edge PHYs is fixed and independent of outline. For each outline, {\workname} uses $L_{\mathrm{die}}/2$ and $W_{\mathrm{die}}/2$ as east/west and north/south access lengths and enumerates lane counts whose length-weighted wire/repeater demand fits the global routing budget; the maximum data-lane count times per-wire rate defines $BW_{\max}$.

\noindent\textbf{Boundary-access and placement check.}
The selected memory-interface and D2D-I/O plans must satisfy the effective die-edge constraint:
\begin{equation}
E_{\mathrm{req}}\le E_{\max}\left(L_{\mathrm{die}},W_{\mathrm{die}};\tau\right)
\end{equation}
where $\tau$ denotes the fixed boundary-placement setting, including the footprint of each boundary-facing block, its legal sides and orientations, the usable interval on each die side, corner keep-outs, reserved edge segments, and minimum inter-block spacing. The scalar inequality is compact notation for an explicit placement-feasibility test rather than a comparison against the aggregate die perimeter. For each unique die-outline and interface-plan combination, {\workname} enumerates the legal side assignments and orientations of the memory-interface and D2D-I/O blocks, and admits the combination only if at least one non-overlapping placement exists within the usable edge intervals.

\vspace{-5pt}
\subsection{Exact Local Structural Reduction}

Direct configuration matching can generate redundant descriptions of the same effective die configuration. Before lifting die configurations to their compatible wafer tilings, {\workname} applies two exact local structural reductions.

\noindent\textbf{Equivalence-class reduction.}
Boundary-constrained blocks may be mirrored or reordered without changing the die footprint, resource totals, placement feasibility, or any evaluator-visible attribute. As illustrated in Fig.~\ref{fig:5-1}, such descriptions belong to the same equivalence class. {\workname} retains one canonical representative from each class.

\noindent\textbf{Same-footprint local strict-dominance reduction.}
Two die configurations may have the same modeled footprint and satisfy the same modeled-feasibility conditions, while one provides no less capability in every compared resource dimension and strictly greater capability in at least one. In Fig.~\ref{fig:5-1}, Designs A and B have the same modeled footprint, but A strictly dominates B in the compared compute, memory, and D2D dimensions.

Within the same die-footprint, compatible-tiling-set, placement, and interface-compatibility class, the evaluators are monotonic in the compared capability dimensions. Therefore, B cannot outperform A and is removed. Configurations with different die geometries, compatible wafer tilings, placement or interface outcomes, or cross-dimensional resource trade-offs are retained.

After reduction, each remaining die configuration is combined with all compatible homogeneous wafer tilings. The resulting distinct modeled-feasible wafer-scale design space is denoted by $\mathcal{D}$ and serves as the input to the subsequent DSE stage.

\vspace{-5pt}

\section{Decision-Domain-Guided Cross-Fidelity Refinement}
\label{sec:hybrid_dse}

This section presents Stage II of {\workname}. Given the distinct modeled-feasible design space $\mathcal{D}$ constructed in Section~IV, the objective is to reduce calls to the designated reference evaluator while retaining the design that would be selected by exhaustive designated-reference evaluation.

The key idea is to replace rank-based winner prediction with bound-based candidate exclusion. A workload- and space-specific disagreement bound maps each low-cost score to a range of reference-consistent outcomes. These ranges induce a \emph{Decision Domain}: the set of candidates that cannot yet be excluded from reference optimality and therefore still require reference evaluation. We first formalize the cross-fidelity disagreement model and then derive the Decision Domain and its conditional containment property.

\begin{figure}[t]
    \setlength{\abovecaptionskip}{0pt}
    \centering
    \includegraphics[width=\columnwidth]{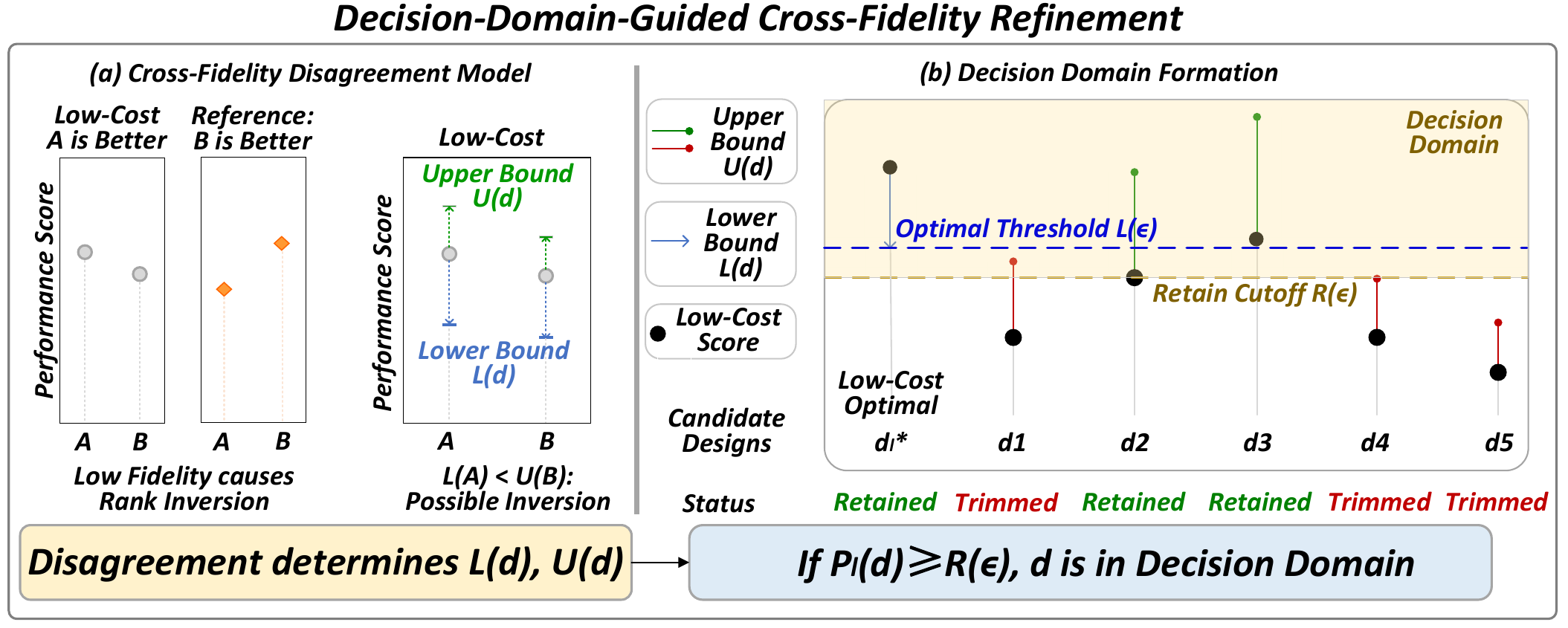}
    \caption{Decision-Domain-guided cross-fidelity refinement. (a) Low-cost and designated-reference evaluators may rank nearby candidates differently; a low-cost-to-reference disagreement bound maps each low-cost score to a reference-consistent performance interval. (b) Given a valid disagreement bound, {\workname} retains candidates whose optimistic reference-performance bound can still reach the pessimistic bound of the low-cost optimum; these candidates form the Decision Domain.}
    \label{fig:6-1}
    \vspace{-17pt}
\end{figure}

\begin{figure*}[t]
    \setlength{\abovecaptionskip}{0pt}
    \centering
    \includegraphics[width=\textwidth]{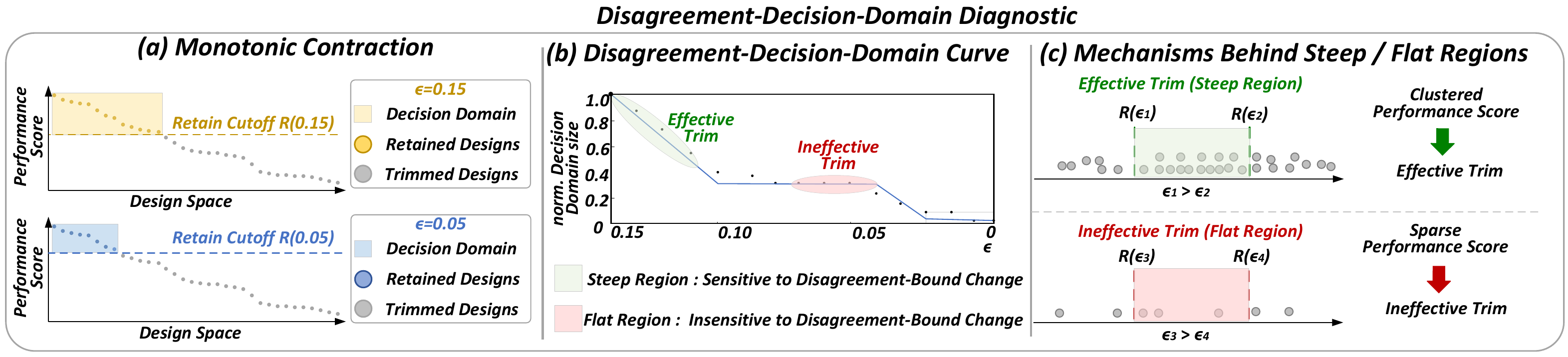}
    \caption{Disagreement--Decision-Domain diagnostic. For a fixed low-cost score distribution, (a) tightening the disagreement bound raises the low-cost-score cutoff and contracts the Decision Domain. (b) The Disagreement--Decision-Domain Curve ({\curvename}) reports the normalized Decision Domain size as the disagreement bound varies. (c) Dense scores near the cutoff produce a sensitive region in which a small reduction in disagreement removes many candidates, whereas sparse scores produce an insensitive region with limited additional contraction.}
    \label{fig:6-2}
    \vspace{-17pt}
\end{figure*}

\vspace{-5pt}

\subsection{Cross-Fidelity Disagreement Model}

For a fixed workload and scalar execution-time objective, the low-cost evaluator covers the complete formed design space $\mathcal{D}$, but its output should not be treated as an exact prediction of designated-reference performance. Instead, {\workname} uses cross-fidelity disagreement to convert each low-cost point estimate into a range of reference-consistent outcomes. For each candidate $d\in\mathcal{D}$, let $P_{\mathrm{L}}(d)$ and $P_{\mathrm{R}}(d)$ denote the performance scores produced by the low-cost and designated reference evaluators, respectively. Each score is defined as the reciprocal of the corresponding execution time, so larger values are better.

We measure the low-cost-to-reference disagreement using the relative performance error
\begin{equation}
e(d)
=
\left|
\frac{P_{\mathrm{L}}(d)}
     {P_{\mathrm{R}}(d)}
-1
\right|,
\qquad
\epsilon
=
\max_{d\in\mathcal{D}} e(d).
\label{eq:cross-fidelity-disagreement}
\end{equation}
Here, $e(d)$ captures the disagreement for one candidate, while $\epsilon$ is the worst-case disagreement over the complete formed design space and is therefore referred to as the full-domain \emph{disagreement bound}.

Under a valid bound $0\leq\epsilon<1$, the designated-reference score of each candidate lies within
\begin{equation}
\frac{P_{\mathrm{L}}(d)}{1+\epsilon}
\leq
P_{\mathrm{R}}(d)
\leq
\frac{P_{\mathrm{L}}(d)}{1-\epsilon},
\qquad
\forall d\in\mathcal{D}.
\label{eq:reference-consistent-interval}
\end{equation}
This interval captures the reference outcomes that remain consistent with the low-cost score and the disagreement bound; it does not assume that the low-cost ranking is preserved.

As illustrated in Fig.~\ref{fig:6-1}(a), the right side maps each low-cost score to its pessimistic and optimistic reference-performance bounds, while the left side shows the resulting ranking ambiguity. Even if $P_{\mathrm{L}}(A)>P_{\mathrm{L}}(B)$, the designated reference evaluator may still produce $P_{\mathrm{R}}(A)<P_{\mathrm{R}}(B)$ when their reference-consistent intervals overlap.

\vspace{-5pt}
\subsection{Decision Domain Formation}

The low-cost optimum provides a natural anchor for candidate exclusion, although it need not be the final selected design. Let $d_{\mathrm{L}}^\star\in\arg\max_{d\in\mathcal{D}}P_{\mathrm{L}}(d)$ denote a low-cost optimum. Because cross-fidelity ranking inversions may occur, $d_{\mathrm{L}}^\star$ may not maximize the designated-reference score. Nevertheless, any designated-reference-optimal candidate must perform at least as well as $d_{\mathrm{L}}^\star$ under reference evaluation.

Under a valid disagreement bound $\epsilon$, Eq.~\eqref{eq:reference-consistent-interval} provides a pessimistic reference-performance bound for $d_{\mathrm{L}}^\star$ and an optimistic reference-performance bound for every other candidate. {\workname} compares these two bounds to determine whether a candidate can be excluded. If even the optimistic reference performance of candidate $\mathcal{d}$ is lower than the pessimistic reference performance of $d_{\mathrm{L}}^\star$, then $\mathcal{d}$ cannot be designated-reference-optimal. Otherwise, its ordering remains unresolved and reference evaluation is still required.

Accordingly, {\workname} defines the retain cutoff $R(\epsilon)$ and the corresponding Decision Domain as
\begin{equation}
\begin{aligned}
R(\epsilon)
&=
\frac{1-\epsilon}{1+\epsilon}
P_{\mathrm{L}}(d_{\mathrm{L}}^\star), \\
\mathcal{C}(\epsilon)
&=
\left\{
d\in\mathcal{D}
\;\middle|\;
P_{\mathrm{L}}(d)\geq R(\epsilon)
\right\}.
\end{aligned}
\label{eq:decision-domain}
\end{equation}
As illustrated in Fig.~\ref{fig:6-1}(b), candidates below $R(\epsilon)$ are trimmed because they cannot reach the reference-performance anchor even under their optimistic bounds. Candidates in $\mathcal{C}(\epsilon)$ remain unresolved under the disagreement bound and are therefore retained for designated-reference evaluation. When candidates are ordered by their low-cost scores, $\mathcal{C}(\epsilon)$ can equivalently be represented as a workload- and space-specific prefix of that ranking. Unlike conventional analytical top-$k$ screening, however, {\workname} does not specify the prefix length or retained fraction in advance; the measured evaluator disagreement determines the retain cutoff, while the low-cost score distribution determines how many candidates lie above it. \textit{Thus, the distinction is not the shape of the retained set, but the disagreement-induced, pair-specific determination of its boundary.}

\noindent\textbf{Conditional containment.}
The Decision Domain provides a conditional containment property. If $\epsilon$ upper-bounds evaluator disagreement over every candidate in $\mathcal{D}$, then every designated-reference optimum is retained. Let $d_{\mathrm{R}}^\star\in\arg\max_{d\in\mathcal{D}}P_{\mathrm{R}}(d)$ denote such an optimum. Reference optimality and the valid reference-consistent intervals imply
\begin{equation}
\frac{P_{\mathrm{L}}(d_{\mathrm{R}}^\star)}{1-\epsilon}
\geq
P_{\mathrm{R}}(d_{\mathrm{R}}^\star)
\geq
P_{\mathrm{R}}(d_{\mathrm{L}}^\star)
\geq
\frac{P_{\mathrm{L}}(d_{\mathrm{L}}^\star)}{1+\epsilon}.
\label{eq:conditional-containment}
\end{equation}
Therefore, $P_{\mathrm{L}}(d_{\mathrm{R}}^\star)\geq R(\epsilon)$, so $d_{\mathrm{R}}^\star$ satisfies the retention condition in Eq.~\eqref{eq:decision-domain} and hence $d_{\mathrm{R}}^\star\in\mathcal{C}(\epsilon)$. This containment property holds whenever $\epsilon$ is valid over the complete formed design space.

\noindent\textbf{Refinement workflow.}
Given a workload- and space-specific disagreement value, {\workname} evaluates the complete formed design space $\mathcal{D}$ with the low-cost evaluator, forms the Decision Domain through Eq.~\eqref{eq:decision-domain}, and applies the designated reference evaluator only to the retained candidates. Under a valid domain-wide bound, returning the highest-scoring candidate in $\mathcal{C}(\epsilon)$ preserves the design selected by exhaustive designated-reference evaluation. If the supplied disagreement value does not permit interval-based exclusion, {\workname} retains the complete space. Section~VI-D describes how the evaluated implementation estimates this input through paired in-domain sampling and accounts for its cost.

\vspace{-5pt}
\subsection{Disagreement-Decision-Domain Diagnostic}

The size of the Decision Domain depends on both the disagreement bound and the distribution of low-cost scores around the resulting retain cutoff. To characterize this relationship, we define an offline diagnostic that keeps the low-cost score vector $\{P_{\mathrm{L}}(d)\mid d\in\mathcal{D}\}$ fixed while varying the disagreement value $\epsilon$.

Equation~\eqref{eq:decision-domain} gives a monotonic relationship. For $0\leq\epsilon_2<\epsilon_1<1$, tightening the disagreement bound raises the retain cutoff and contracts the Decision Domain:
\[
R(\epsilon_2)>R(\epsilon_1),
\qquad
\mathcal{C}(\epsilon_2)\subseteq\mathcal{C}(\epsilon_1).
\]
Thus, reducing the disagreement can only preserve or contract the Decision Domain, as illustrated in Fig.~\ref{fig:6-2}(a). We summarize this behavior by plotting the normalized domain size $|\mathcal{C}(\epsilon)|/|\mathcal{D}|$ as $\epsilon$ varies, yielding the \emph{Disagreement--Decision-Domain Curve} ({\curvename}) shown in Fig.~\ref{fig:6-2}(b).

The amount of contraction depends on the density of low-cost scores near the retain cutoff. As illustrated in Fig.~\ref{fig:6-2}(c), if many candidates lie between the cutoffs associated with two disagreement values, a small tightening removes many candidates and produces a steep curve segment. If few candidates lie in this interval, the same tightening produces a flat segment with limited additional contraction. The {\curvename} is an offline diagnostic rather than an additional step in the {\workname} workflow. It characterizes how evaluator disagreement and the score distribution jointly determine the retained-domain size and the resulting demand for designated-reference evaluation.

\vspace{-5pt}
\section{Evaluation}
\label{sec:eval}

\begin{table}[b]
    \vspace{-15pt}
    \centering
    \caption{LLM-training workloads and execution settings.}

    \label{tab:workloads}
    \setlength{\tabcolsep}{1.0pt}
    \renewcommand{\arraystretch}{1.05}
    \scriptsize
    \begin{tabular}{@{}c l r r r r c c c c@{}}
        \toprule
        \textbf{ID} &
        \textbf{Model} &
        \textbf{MB} &
        \textbf{Seq.} &
        \textbf{GA} &
        \textbf{CKPT} &
        \textbf{Opt.} &
        \shortstack{\textbf{Wgt.}\\\textbf{Prec.}} &
        \shortstack{\textbf{State}\\\textbf{Prec.}} &
        \textbf{Parallel} \\
        \midrule
        0 & Llama-3B    & 512 & 1,024 & 8  & 4 & AdamW & 16-bit & 32-bit & DP+TP \\
        1 & Llama-3B    & 32  & 512   & 1  & 4 & AdamW & 16-bit & 32-bit & DP+TP \\
        2 & Llama-8B    & 256 & 2,048 & 16 & 4 & AdamW & 16-bit & 32-bit & DP+TP \\
        3 & Qwen2.5-32B & 64  & 2,048 & 16 & 4 & AdamW & 16-bit & 32-bit & DP+TP \\
        4 & Llama-70B   & 32  & 2,048 & 16 & 4 & AdamW & 16-bit & 32-bit & DP+TP \\
        5 & Llama-70B   & 8   & 8,192 & 16 & 2 & AdamW & 16-bit & 32-bit & DP+TP \\
        6 & Llama-405B  & 4   & 2,048 & 32 & 2 & AdamW & 16-bit & 32-bit & DP+TP \\
        \bottomrule
    \end{tabular}
\end{table}

\begin{table}[b]
    \vspace{-18pt}
    \centering
    \caption{Reference-verifiable design-space configurations.}

    \label{tab:design_space_configuration}
    \setlength{\tabcolsep}{1.5pt}
    \renewcommand{\arraystretch}{1.05}
    \scriptsize
    \begin{tabular}{@{}c c c c c c c c r@{}}
        \toprule
        \multirow{2}{*}{\textbf{DS}} &
        \multicolumn{2}{c}{\shortstack{\textbf{Die-Outline Range}\\\textbf{(mm)}}} &
        \multirow{2}{*}{\shortstack{\textbf{Comp.}\\\textbf{Plans}}} &
        \multirow{2}{*}{\shortstack{\textbf{Mem.}\\\textbf{Plans}}} &
        \multirow{2}{*}{\shortstack{\textbf{D2D}\\\textbf{Plans}}} &
        \multicolumn{2}{c}{\shortstack{\textbf{Compatible}\\\textbf{Tiling}}} &
        \multirow{2}{*}{\(\boldsymbol{|\mathcal{D}|}\)} \\
        \cmidrule(lr){2-3}
        \cmidrule(lr){7-8}
        &
        \textbf{Length} &
        \textbf{Width} &
        &
        &
        &
        \textbf{Rows} &
        \textbf{Cols.} &
        \\
        \midrule
        a & 8--33  & 8--26  & 16 & 18 & 2 & 6--27 & 8--27 & 192   \\
        b & 10--33 & 10--26 & 12 & 27 & 3 & 6--21 & 8--21 & 272   \\
        c & 14--33 & 14--26 & 15 & 25 & 4 & 6--15 & 8--15 & 405   \\
        d & 10--33 & 10--26 & 40 & 9  & 4 & 6--21 & 8--21 & 5,655 \\
        e & 9--33  & 9--26  & 8  & 32 & 3 & 6--24 & 8--24 & 351   \\
        f & 12--33 & 12--26 & 12 & 30 & 2 & 6--18 & 8--18 & 248   \\
        g & 8--33  & 8--26  & 32 & 12 & 2 & 6--27 & 8--27 & 603   \\
        h & 12--33 & 12--26 & 24 & 16 & 1 & 6--18 & 8--18 & 179   \\
        i & 10--33 & 10--26 & 24 & 16 & 8 & 6--21 & 8--21 & 372   \\
        j & 8--33  & 8--26  & 32 & 18 & 4 & 6--27 & 8--27 & 677   \\
        \bottomrule
    \end{tabular}
\end{table}

\vspace{-5pt}
\subsection{Experimental Setup}

Tables~\ref{tab:workloads} and~\ref{tab:design_space_configuration} summarize seven LLM-training workloads and ten reference-verifiable design spaces, yielding 70 workload--design-space pairs. In Table~\ref{tab:workloads}, MB, Seq., GA, and CKPT denote wafer microbatch size, sequence length, gradient-accumulation steps, and checkpoint interval; Wgt. Prec. and Opt.-State Prec. denote weight and optimizer-state precision; and DP and TP denote data and tensor parallelism. All spaces assume a 5\,nm process, informed by published 5\,nm logic and SRAM technology characteristics~\cite{Yeap2019N5,Chang2021N5SRAM}, a \(220\,\mathrm{mm}\times220\,\mathrm{mm}\) usable wafer region, and a \(33\,\mathrm{mm}\times26\,\mathrm{mm}\) reticle limit. We instantiate $\tau_{\mathrm{D2D}}$ using a literature-derived 5\,nm short-reach PHY that provides \(25.2\,\mathrm{Gb/s}\) per direction per physical wire at a characterized \(1.2\,\mathrm{mm}\) channel~\cite{NishiJSSC2023}. Following the BoW organization, 16 data wires per PHY slice provide \(403.2\,\mathrm{Gb/s}\) per direction~\cite{BoWPHY}. For footprint accounting, we use the BoW advanced-package reference organization at a \(40\,\mu\mathrm{m}\) bump pitch, whose 16-slice arrangement occupies \(1.60\,\mathrm{mm}\) along the die edge and \(0.42\,\mathrm{mm}\) in edge-normal depth. We instantiate $\tau_{\mathrm{edge}}$ with a \(0.20\,\mathrm{mm}\) keep-out at each corner, a centered reserved segment covering 10\% of each die side, and a minimum inter-block spacing of \(20\,\mu\mathrm{m}\). Performance Score is the reciprocal of workload execution time. Additional spaces containing up to 46,782 candidates are used only for the construction-scaling study.

All experiments run on a single-socket x86-64 server with a six-core Intel Core i7-6850K processor at a 3.60\,GHz base frequency and 128\,GB of memory, executing up to 12 simulator jobs concurrently. The exhaustive reference corpus is generated once and reused across all randomized search-policy runs.

Chakra~\cite{Chakra} generates the training traces, which are evaluated by ASTRA-sim's analytical backend and the designated ASTRA-sim+ns-3 reference backend~\cite{ASTRA-sim2.0,ns-3}. Exhaustive ASTRA-sim+ns-3 evaluation, denoted Exhaustive Reference, establishes the designated-reference optimum for each of the 70 pairs. These results are used only for post-search validation and remain unavailable to all DSE methods during exploration.

For end-to-end search efficiency, we compare {\workname} with Exhaustive Reference. For selected-design quality, we compare Exhaustive Analytical, {\workname}, Simulated Annealing (SA)~\cite{SimulatedAnnealing}, Theseus~\cite{Theseus}, and Polaris~\cite{Polaris}, using Exhaustive Reference as the designated-reference optimum. Where applicable, all methods use ASTRA-sim analytical and ASTRA-sim+ns-3 as the common low- and high-fidelity evaluators, optimize the same execution-time objective, and search the same formed design spaces. We additionally cross-validate ASTRA-sim+ns-3 against gem5 Garnet 3.0~\cite{gem5,Garnet,heterogarnet}.

\vspace{-5pt}
\subsection{Cross-Validation of the Designated Reference Backend}

ASTRA-sim+ns-3 serves as the designated reference backend in our main evaluation rather than silicon ground truth. To examine whether the resulting architecture decisions depend on ns-3-specific modeling choices, we cross-validate four design-space/workload settings using gem5 Garnet 3.0~\cite{gem5,Garnet,heterogarnet}. Garnet follows an implementation path independent of ns-3 and models router pipelines, input buffering, virtual channels, switch arbitration, credit-based flow control, and link traversal at cycle level.

For each evaluated wafer design, Garnet uses the same die-array topology, routing policy, link bandwidth, and link latency as ASTRA-sim+ns-3, and replays the same workload communication events as packet injections. Compute and memory timing remain unchanged, so this experiment isolates whether the interconnect model changes the resulting architecture ranking and selected design. The evaluated settings cover design spaces a and j and two workloads with distinct communication characteristics: W1, the communication-intensive Llama-3B workload, and W6, the Llama-405B workload.

\begin{table}[t]
    \centering
    \caption{Decision-level cross-validation against gem5 Garnet.}
    \vspace{-7pt}
    \label{tab:reference-cross-validation}
    \setlength{\tabcolsep}{5.0pt}
    \renewcommand{\arraystretch}{1.05}
    \scriptsize
    \begin{tabular}{@{}l c c c@{}}
        \toprule
        \textbf{DS / Workload} &
        \textbf{\# Designs} &
        \textbf{Spearman $\rho$} &
        \textbf{Same Best} \\
        \midrule
        a / W1 & 192 & 0.9901 & Yes \\
        a / W6 & 192 & 0.9984 & Yes \\
        j / W1 & 677 & 0.9906 & Yes \\
        j / W6 & 677 & 0.9993 & Yes \\
        \bottomrule
    \end{tabular}
    \vspace{-15pt}
\end{table}

Across the four settings, ASTRA-sim+ns-3 and Garnet achieve a mean Spearman rank correlation of 0.9946 and select the same best design in every case. This result supports decision-level consistency between the two interconnect models for the evaluated settings.

\vspace{-5pt}
\subsection{Physical-Formulation Effectiveness and Large-Space Construction}

We evaluate Stage~I from two perspectives: how much the formulation contracts the construction space, and whether the physical-implication checks exclude candidates that appear competitive under low-cost evaluation. For each of the ten reference-verifiable design spaces, we compare three nested spaces. The \emph{Area-Feasible} space satisfies the reticle, compatible-tiling, and aggregate-area constraints. The \emph{Physical-Implication-Aware} space further enforces the die-size-dependent D2D-capability, die-edge-access, and placement constraints. The final {\workname} space additionally applies equivalence-class canonicalization and same-footprint local strict-dominance reduction.

To measure whether the physical checks affect performance-relevant candidates, we define the Top-10\% Physical-Rejection Rate. For each workload, we rank all Area-Feasible candidates using the analytical evaluator, take the highest-scoring 10\%, and measure the fraction removed by the complete physical-implication checks. The rate is computed separately for each workload and then averaged across the seven workloads for each design space.

\begin{figure}[t]
    \setlength{\abovecaptionskip}{0pt}
    \centering
    \includegraphics[width=\columnwidth]{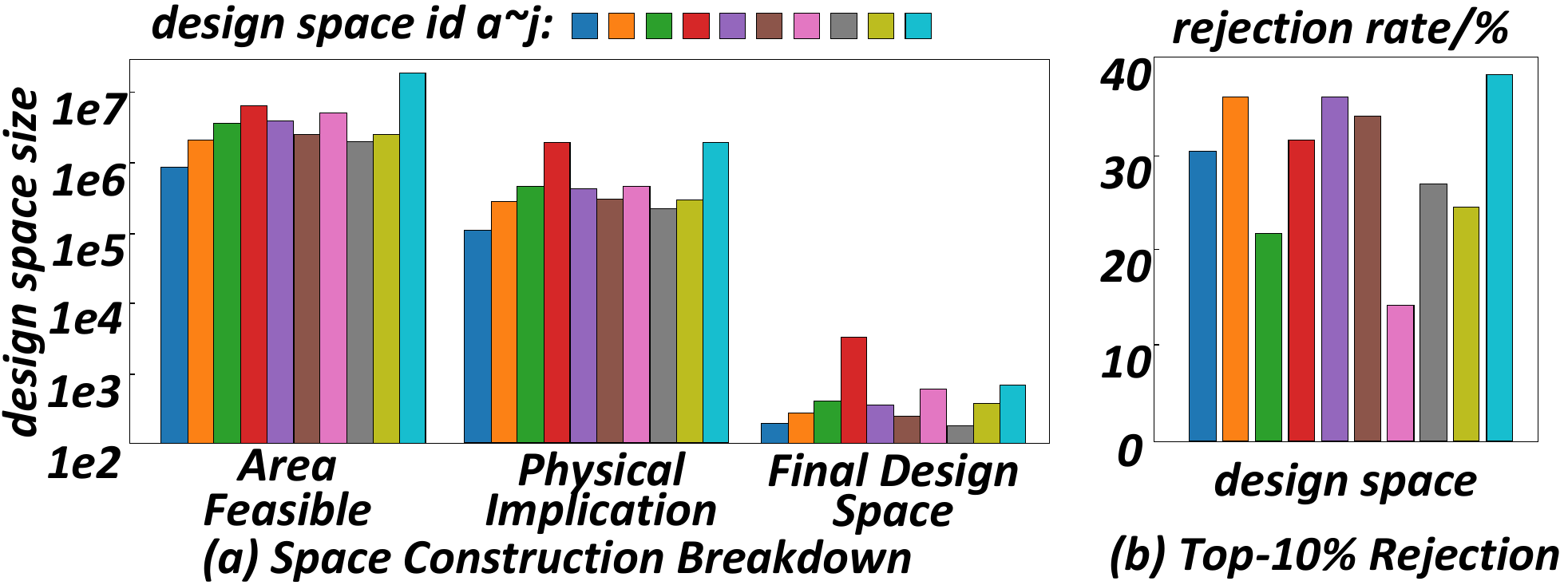}
    \caption{Physical-formulation ablation across the ten reference-verifiable design spaces. (a) Candidate counts in the Area-Feasible, Physical-Implication-Aware, and final {\workname} spaces. (b) Mean Top-10\% Physical-Rejection Rate across the seven workloads for each design space.}
    \label{fig:7-3}
    \vspace{-15pt}
\end{figure}

As shown in Fig.~\ref{fig:7-3}(a), the complete physical-implication checks remove 69.4\%--90.7\% of the Area-Feasible candidates across the ten design spaces, with an overall mean of 86.4\%. The subsequent exact local reductions further eliminate 99.83\%--99.96\% of the Physical-Implication-Aware candidates, with an overall mean of 99.89\%.

More importantly, Fig.~\ref{fig:7-3}(b) shows mean Top-10\% Physical-Rejection Rates ranging from 14.1\% to 38.2\%, with an overall mean of 29.4\%. Thus, aggregate-area feasibility does not merely admit low-ranked tail candidates; it also admits analytically competitive designs that violate the modeled D2D, die-edge, or placement requirements. For example, in design space~f, the analytical-best Area-Feasible candidate is excluded because it violates the D2D-capability constraint.

\begin{figure*}[t]
    \setlength{\abovecaptionskip}{0pt}
    \centering
    \includegraphics[width=\textwidth]{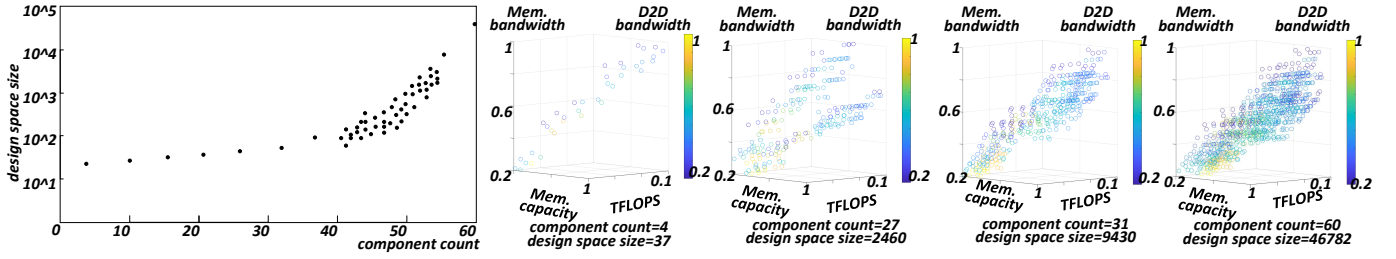}
    \caption{Stage-I construction scale and structure of the modeled-feasible design spaces. The left panel shows the final design-space size as component-library diversity increases. The remaining panels show representative spaces containing 37, 2,460, 9,430, and 46,782 distinct candidates. The axes denote system TFLOPS, memory capacity, and memory bandwidth, while color encodes D2D bandwidth. These larger spaces extend the construction study beyond the ten reference-verifiable spaces used for end-to-end evaluation.}
    \label{fig:7-4}
    \vspace{-15pt}
\end{figure*}

Figure~\ref{fig:7-4} evaluates Stage-I construction over a broader sweep of component-library configurations. As component-library diversity increases, the resulting distinct modeled-feasible space grows to 46,782 candidates in the largest evaluated configuration. The representative spaces remain discrete and irregular across compute throughput, memory capacity, memory bandwidth, and D2D bandwidth, showing that the formulation preserves diverse cross-dimensional trade-offs rather than collapsing the space into a narrow family of similar configurations.

\vspace{-5pt}
\subsection{Practical Calibration}

To instantiate the workload- and space-specific disagreement value required by the refinement stage, the evaluated implementation uses uniform paired sampling within each workload/design-space pair. After evaluating the complete space with the low-cost evaluator, we uniformly sample $S\subseteq\mathcal{D}$ with $|S|=\lceil\rho|\mathcal{D}|\rceil$, additionally evaluate these candidates using the designated reference evaluator, and compute $\hat{\epsilon}=\max_{d\in S}e(d)$. 

Because $\hat{\epsilon}$ is a sampled maximum, it may underestimate the full-domain disagreement bound. Full-domain disagreement coverage is sufficient for the conditional containment result in Section~V-B, but is not necessary for retaining the reference optimum in a particular workload/design-space pair: the largest pointwise disagreement may occur on a candidate far below the retain cutoff and therefore leave the reference-optimum retention decision unchanged. We therefore treat $\hat{\epsilon}$ as a practical disagreement estimate and select the sampling rate $\rho$ empirically through leave-one-design-space-out validation. For each held-out design space, we evaluate sampling rates from 1\% to 15\% on the other nine spaces, select the smallest rate with no observed recovery failure, and test that rate on the held-out space without further tuning. The fold-selected rates range from 8\% to 10\%; we therefore use 10\%, the maximum selected rate, as a conservative common operating point.

Section~VI-F evaluates the resulting selected-design quality over 20 independent calibration draws per workload/design-space pair. In a retrospective diagnostic, the largest observed underestimation of the full-domain disagreement is 1.47 percentage points. The designated-reference results obtained during calibration are cached and reused: after forming $\mathcal{C}(\hat{\epsilon})$, only candidates in $\mathcal{C}(\hat{\epsilon})\setminus S$ require additional designated-reference evaluation, while final selection considers all reference-evaluated candidates in $\mathcal{C}(\hat{\epsilon})\cup S$. Calibration evaluations are included in the total reference cost reported in the end-to-end results.

\begin{figure}[t]
    \setlength{\abovecaptionskip}{0pt}
    \centering
    \includegraphics[width=\columnwidth]{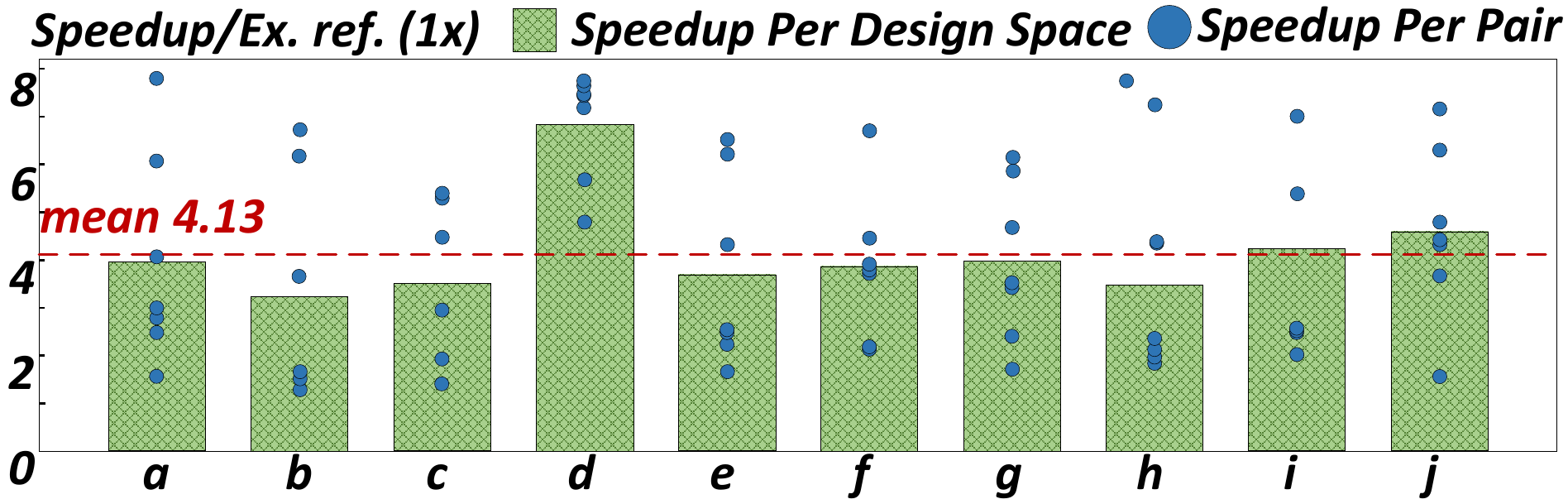}
    \caption{End-to-end {\workname} speedup over Exhaustive Reference. Each bar reports the mean across the seven workloads for one design space; the dashed line indicates the overall mean across all 70 workload/design-space pairs.}
    \label{fig:7-5}
    \vspace{-15pt}
\end{figure}

\begin{table}[b]
    \vspace{-15pt}
    \centering
    \caption{Decision-Domain and reference-evaluation fractions.}
    \vspace{-7pt}
    \label{tab:decision-domain-statistics}
    \setlength{\tabcolsep}{5.0pt}
    \renewcommand{\arraystretch}{1.05}
    \scriptsize
    \begin{tabular}{@{}lrrrr@{}}
        \toprule
        \textbf{Metric} &
        \textbf{Mean} &
        \textbf{Median} &
        \textbf{95th Pctl.} &
        \textbf{Maximum} \\
        \midrule
        Decision-Domain fraction &
        13.58\% & 11.94\% & 33.98\% & 50.37\% \\
        Total reference fraction &
        20.42\% & 19.24\% & 38.41\% & 52.94\% \\
        \bottomrule
    \end{tabular}
\end{table}

\vspace{-5pt}
\subsection{End-to-End Search Efficiency}

For each workload/design-space pair, {\workname}'s end-to-end runtime includes full-space analytical evaluation, 10\% paired calibration, and all additional designated-reference evaluations required by the resulting Decision Domain. Calibration results are cached and reused when sampled candidates also belong to the retained domain. The Decision-Domain fraction is normalized to the formed design-space size, while the total reference fraction counts all distinct candidates evaluated by the designated reference backend during either calibration or Decision-Domain evaluation. We report speedup relative to Exhaustive Reference over the same formed design space.

As shown in Table~\ref{tab:decision-domain-statistics}, the resulting Decision Domain contains 13.58\% of the formed design space on average and 11.94\% at the median. The 95th-percentile and maximum fractions are 33.98\% and 50.37\%, respectively. After including calibration samples outside the retained domain, {\workname} evaluates 20.42\% of the candidates with the designated reference backend on average, with a median of 19.24\% and a maximum of 52.94\%.

Figure~\ref{fig:7-5} reports the resulting end-to-end speedup across the 70 workload/design-space pairs. After accounting for full-space analytical evaluation, calibration, additional designated-reference evaluation, and algorithmic overhead, {\workname}'s average runtime is 24.21\% of Exhaustive Reference, corresponding to an average speedup of 4.13$\times$ and a maximum speedup of 7.80$\times$. Variation across pairs largely reflects the retained-domain size and the candidate-level cost of designated-reference evaluation: smaller retained domains generally require fewer additional reference evaluations and achieve greater speedup.

\vspace{-5pt}
\subsection{Selected-Design Quality}
\begin{table}[t]
\vspace{-7pt}
    \centering
    \caption{Run-level selected-design quality.}
    \vspace{-7pt}
    \label{tab:selected-design-summary}
    \setlength{\tabcolsep}{2.0pt}
    \renewcommand{\arraystretch}{1.05}
    \scriptsize
    \begin{tabular}{@{}lcccc@{}}
        \toprule
        \textbf{Method} &
        \shortstack{\textbf{Exact}\\\textbf{Recovery}} &
        \shortstack{\textbf{Mean Norm.}\\\textbf{Perf.}} &
        \shortstack{\textbf{5th-Pctl.}\\\textbf{Norm. Perf.}} &
        \shortstack{\textbf{Worst Norm.}\\\textbf{Perf.}} \\
        \midrule
        {\workname} &
        \textbf{100\% (1400/1400)} &
        \textbf{1.000} &
        \textbf{1.000} &
        \textbf{1.000} \\
        Polaris &
        84.86\% (1188/1400) &
        0.9959 &
        0.9921 &
        0.6721 \\
        Theseus &
        25.50\% (357/1400) &
        0.9691 &
        0.8457 &
        0.0056 \\
        SA &
        24.14\% (338/1400) &
        0.9546 &
        0.7874 &
        0.0055 \\
        Exhaustive Analytical &
        7.14\% (5/70) &
        0.9329 &
        0.8182 &
        0.7325 \\
        \bottomrule
    \end{tabular}
    \vspace{-15pt}
\end{table}
Exhaustive Analytical selects the low-cost optimum, while Exhaustive Reference provides the designated-reference optimum used for normalization. In the evaluated implementation, {\workname} uses 10\% paired calibration followed by designated-reference evaluation of the resulting Decision Domain. Theseus and Polaris retain their respective multi-fidelity procedures using the common analytical and designated-reference backends. SA follows its standard search procedure under the same execution-time objective, formed design space, and runtime accounting. This common setup isolates the methods' evaluation-allocation and design-selection policies under the same objective, candidate space, and evaluator pair.

{\workname} is an adaptive-cost method: its realized reference-evaluation cost is determined by the workload- and space-specific disagreement and the resulting Decision-Domain size. In contrast, SA, Theseus, and Polaris require an ex ante stopping budget. We therefore assign these methods a fixed per-pair runtime budget equal to 25\% of Exhaustive Reference, selected before any search run. This budget slightly exceeds {\workname}'s measured mean end-to-end runtime of 24.21\% of Exhaustive Reference. The comparison therefore evaluates adaptive-cost refinement against fixed-budget search at the same benchmark-level cost scale; it does not impose hindsight pairwise matching to {\workname}'s realized cost, which is unavailable to a competing method before execution. Each method with a stochastic procedure is evaluated over 20 independent runs per workload/design-space pair. All analytical evaluations, designated-reference evaluations, and algorithmic overheads are included when enforcing the runtime budget.

As shown in Table~\ref{tab:selected-design-summary}, {\workname} recovers the exhaustive designated-reference optimum in all 1,400 runs across the 70 workload/design-space pairs and 20 independent calibration draws per pair; its mean, 5th-percentile, and worst-case normalized performance are therefore all 1.000. Polaris provides the strongest baseline average and lower-tail performance, but recovers the exact optimum in only 84.86\% of its runs and reaches a worst-case normalized performance of 0.6721. Theseus and SA recover the exact optimum in only 25.50\% and 24.14\% of their runs, respectively, and exhibit rare near-zero tail outcomes. Exhaustive Analytical is deterministic and recovers the designated-reference optimum in only 5 of the 70 pairs. These results show that high average performance can conceal substantial failures for particular workloads or random runs. Despite receiving a slightly larger runtime budget than {\workname} on average, none of the baselines matches its exact-recovery rate or worst-case performance.

\vspace{-5pt}
\subsection{Workload-Dependent Architectural Choices}

\begin{figure}[t]
    \setlength{\abovecaptionskip}{0pt}
    \centering
    \includegraphics[width=\columnwidth]{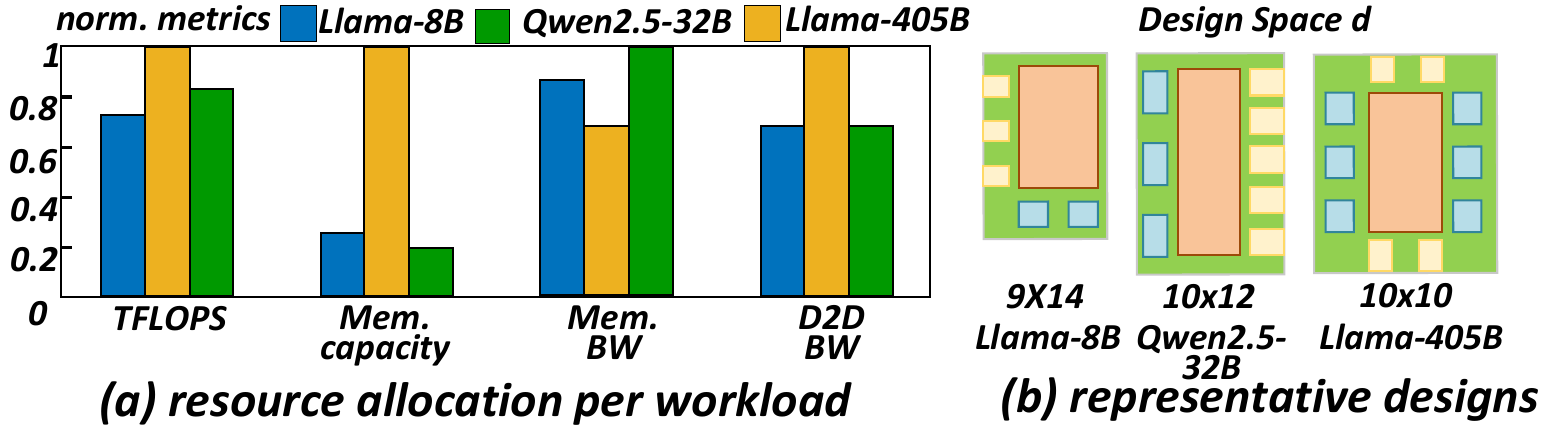}
    \caption{Workload-dependent reference-optimal designs. (a) Normalized compute throughput, memory capacity, memory bandwidth, and D2D bandwidth. (b) Representative die-array organizations.}
    \label{fig:7-7}
    \vspace{-18pt}
\end{figure}

Fig.~\ref{fig:7-7} compares the reference-optimal designs for three representative LLM-training workloads. Resource metrics are normalized to the maximum among the shown designs.

The preferred resource balance and die-array organization vary across workloads: the representative optima use $9\times14$, $10\times12$, and $10\times10$ organizations and allocate substantially different proportions of compute, memory, and D2D capability. These results confirm that no single candidate is reference-optimal across the evaluated workloads, motivating workload-specific wafer-scale DSE.

\vspace{-5pt}
\section{Related Work}

\noindent\textbf{Industrial and multi-stage DSE practice.} Modern chip development uses workload-driven modeling, architecture exploration, implementation sweeps, and expert-guided trade-off analysis, often under names such as pathfinding, DTCO, or STCO~\cite{IBMFullSystemSimulation2006,GoogleApollo2021,TSMCDTCO2022,RochaSTCO2024}. These flows typically screen broad spaces with low-cost models and reserve detailed simulation or implementation feedback for fewer candidates. However, publicly documented practices provide limited assurance that such screening retains the design preferred by a designated reference evaluator. {\workname} addresses this selection problem without assuming that industrial exploration is either purely manual or fully automated.

\noindent\textbf{Wafer-scale exploration and physical feasibility.} Commercial systems, physical-design studies, communication fabrics, and application demonstrations expose the coupled effects of compute, memory, interconnect, redundancy, mapping, and physical organization in wafer-scale systems~\cite{Wafer-Scale,Cerebras,Dojo,Scale-out-processor,SoWX2025,FRED2025,James2020PhysicalMapping,Rocki2020Stencil,TMAC,WSC-LLM,WATOS,Theseus}. Physical-mapping work studies placement on wafer-scale accelerators, while SoW-X and FRED examine system-level integration and collective-oriented on-wafer fabrics. TMAC, WSC-LLM, and WATOS co-explore wafer-scale architecture choices with workload mapping, serving, or training strategies, whereas Theseus is the closest general wafer-scale DSE framework, combining system and physical constraints with multi-fidelity multi-objective Bayesian optimization using analytical and learned GNN-based evaluators. Such approaches search parameterized spaces for high-quality or Pareto-efficient designs. In contrast, within its homogeneous repeated-die scope, {\workname} constructs the complete distinct modeled-feasible domain through physical-implication mapping, legality-aware matching, and exact local reduction before performance exploration. The distinction is not that prior work ignores physical constraints, but that {\workname} makes complete modeled-feasible-domain construction a first-class stage.

\noindent\textbf{Wafer-scale modeling and workload orchestration.} Complementary work optimizes the execution stack for tiled or wafer-scale substrates. PALM provides event-driven performance modeling for large tiled accelerators~\cite{PALM}; TEMP develops memory-efficient, topology-aware tensor partitioning and mapping~\cite{TEMP}; and ReThermal co-designs compile-time and runtime scheduling under wafer-scale thermal constraints~\cite{ReThermal}. Spatial-Aware Orchestration and MOCAP target attention placement and long-context prefill pipelining, respectively~\cite{SpatialAttention,MOCAP}, while STAR combines sparse-attention algorithm--hardware co-design with deployment on a multi-core spatial architecture~\cite{STAR}. These methods optimize particular mappings, schedules, simulators, or workload-specific architectures and are complementary to {\workname}: they can serve as workload policies, modeled constraints, architectural components, or evaluation backends within the proposed exploration methodology.

\noindent\textbf{Architecture and training simulation.} A broader simulator ecosystem models DNN dataflows, spatial-accelerator mappings, multi-chip manycore systems, and distributed model training at different abstraction levels. MAESTRO and Union support data-centric mapping analysis and HW--SW co-design for spatial accelerators~\cite{Kwon2020MAESTRO,Jeong2021Union}; MuchiSim targets design exploration of multi-chip manycore systems~\cite{OrenesVera2024MuchiSim}; and Proteus and vTrain model distributed DNN and LLM training performance~\cite{Duan2024Proteus,Bang2024vTrain}.

\noindent\textbf{Multi-fidelity hardware DSE.} Prior work applies multi-fidelity optimization to HLS directives~\cite{Lo2018MultiFidelityHLS,Sun2021CorrelatedMFOHLS}, microarchitecture search~\cite{Fan2024FNNMFRL,Liu2025SwiftOrExact}, accelerator co-design~\cite{Polaris}, chiplet exploration~\cite{Chen2026CHASE}, and FPGA network-switch co-design~\cite{Li2026SPAC}. For example, Polaris transfers a low-fidelity representation trained with Timeloop~\cite{Timeloop} to a Starlight predictor using selected FireSim RTL samples~\cite{FireSim}, and embeds it in Bayesian optimization. These methods are primarily search-centric: uncertainty estimates, predicted rankings, or domain-specific constraints direct evaluation toward promising candidates, while many candidates may receive neither low-cost nor designated-reference evaluation. {\workname} is instead confirmation-centric: it evaluates the complete modeled-feasible space with the low-cost backend, uses workload- and space-specific evaluator disagreement to induce the Decision Domain, and applies designated-reference evaluation to every retained candidate. Under a valid full-domain disagreement bound, this domain contains the designated-reference optimum; the evaluated implementation estimates disagreement through paired in-domain sampling and validates end-to-end recovery empirically.

\vspace{-5pt}
\section{Conclusion}
This paper presented {\workname}, a reusable methodology for workload-specific wafer-scale DSE. {\workname} combines physical-implication-aware design-space formulation with Decision-Domain-guided cross-fidelity refinement to construct distinct modeled-feasible spaces and focus designated-reference evaluation on candidates that may remain optimal. Under a valid domain-wide disagreement bound, the Decision Domain contains the designated-reference optimum.


\bibliographystyle{IEEEtranS}
\bibliography{refs}

@inproceedings{ASTRA-SIM,
  author    = {Rashidi, Saeed and Sridharan, Srinivas and Srinivasan, Sudarshan and Krishna, Tushar},
  title     = {{ASTRA-sim}: Enabling {SW/HW} Co-Design Exploration for Distributed {DL} Training Platforms},
  booktitle = {2020 IEEE International Symposium on Performance Analysis of Systems and Software (ISPASS)},
  year      = {2020},
  pages     = {81--92},
  doi       = {10.1109/ISPASS48437.2020.00018}
}

@inproceedings{ASTRA-sim2.0,
  author    = {Won, William and Heo, Taekyung and Rashidi, Saeed and Sridharan, Srinivas and Srinivasan, Sudarshan and Krishna, Tushar},
  title     = {{ASTRA-sim2.0}: Modeling Hierarchical Networks and Disaggregated Systems for Large-Model Training at Scale},
  booktitle = {2023 IEEE International Symposium on Performance Analysis of Systems and Software (ISPASS)},
  year      = {2023},
  pages     = {283--294},
  doi       = {10.1109/ISPASS57527.2023.00035}
}

@misc{BoWPHY,
  author={{Open Compute Project ODSA BoW Workstream}},
  title={Bunch of Wires (BoW) PHY Specification},
  year={2023},
  note={Draft Version 1.9d}
}

@article{Cerebras,
  author  = {Lauterbach, Gary},
  title   = {The Path to Successful Wafer-Scale Integration: The {Cerebras} Story},
  journal = {IEEE Micro},
  year    = {2021},
  volume  = {41},
  number  = {6},
  pages   = {52--57},
  doi     = {10.1109/MM.2021.3112025}
}

@article{Chakra,
  author        = {Sridharan, Srinivas and Balogh, Andy and Beckmann, Bradford M. and Coutinho, Brian and Feng, Louis and Fu, Sheng and Gao, Sanshan and Garakani, Mehryar and Heo, Taekyung and Kanter, David and Ladd, Josh and Li, Ziwei and Liu, Winston and Man, Changhai and Mihailescu, Dan and More, Spandan and Park, Joongun and Ramachandran, Ashwin and Ramakrishnaiah, Vinay and Rashidi, Saeed and Reddi, Vijay Janapa and Sharma, Puneet and Tian, Phio and Won, William and Wu, Hanjiang and Xu, Huan and Yoo, Jinsun and Krishna, Tushar},
  title         = {{MLCommons Chakra}: Advancing Performance Benchmarking and Co-Design Using Standardized Execution Traces},
  year          = {2026},
  journal       = {arXiv preprint arXiv:2605.11333},
  url           = {https://arxiv.org/abs/2605.11333}
}

@inproceedings{Chen2026CHASE,
  author    = {Chen, Shixin and Zhang, Hengyuan and Zhai, Jianwang and Yu, Bei},
  title     = {{CHASE}: A Chiplet Architecture Simulation and Exploration Framework with Decoupled Multi-Fidelity Optimization},
  booktitle = {Proceedings of the 2026 International Symposium on Physical Design},
  year      = {2026},
  pages     = {153--161},
  publisher = {ACM},
  doi       = {10.1145/3764386.3779577}
}

@article{Dojo,
  author  = {Talpes, Emil and Das Sarma, Debjit and Williams, Doug and Arora, Sahil and Kunjan, Thomas and Floering, Benjamin and Jalote, Ankit and Hsiong, Christopher and Poorna, Chandrasekhar and Samant, Vaidehi and Sicilia, John and Nivarti, Anantha Kumar and Ramachandran, Raghuvir and Fischer, Tim and Herzberg, Ben and McGee, Bill and Venkataramanan, Ganesh and Banon, Pete},
  title   = {The Microarchitecture of {Dojo}, {Tesla}'s Exa-Scale Computer},
  journal = {IEEE Micro},
  year    = {2023},
  volume  = {43},
  number  = {3},
  pages   = {31--39},
  doi     = {10.1109/MM.2023.3258906}
}

@inproceedings{Fan2024FNNMFRL,
  author    = {Fan, Hanwei and Wang, Ya and Li, Sicheng and Liang, Tingyuan and Zhang, Wei},
  title     = {Explainable Fuzzy Neural Network with Multi-Fidelity Reinforcement Learning for Micro-Architecture Design Space Exploration},
  booktitle = {Proceedings of the 61st ACM/IEEE Design Automation Conference},
  year      = {2024},
  articleno = {7},
  pages     = {7:1--7:6},
  publisher = {ACM},
  doi       = {10.1145/3649329.3657350}
}

@inproceedings{FireSim,
  author    = {Karandikar, Sagar and Mao, Howard and Kim, Donggyu and Biancolin, David and Amid, Alon and Lee, Dayeol and Pemberton, Nathan and Amaro, Emmanuel and Schmidt, Colin and Chopra, Aditya and Huang, Qijing and Kovacs, Kyle and Nikolic, Borivoje and Katz, Randy H. and Bachrach, Jonathan and Asanovic, Krste},
  title     = {{FireSim}: {FPGA}-Accelerated Cycle-Exact Scale-Out System Simulation in the Public Cloud},
  booktitle = {2018 ACM/IEEE 45th Annual International Symposium on Computer Architecture (ISCA)},
  year      = {2018},
  pages     = {29--42},
  doi       = {10.1109/ISCA.2018.00014}
}

@INPROCEEDINGS{Garnet,
  author={Agarwal, Niket and Krishna, Tushar and Peh, Li-Shiuan and Jha, Niraj K.},
  booktitle={2009 IEEE International Symposium on Performance Analysis of Systems and Software}, 
  title={GARNET: A detailed on-chip network model inside a full-system simulator}, 
  year={2009},
  volume={},
  number={},
  pages={33-42},
  doi={10.1109/ISPASS.2009.4919636}}

@inproceedings{GoogleApollo2021,
  author    = {Yazdanbakhsh, Amir and Angermueller, Christof and Akin, Berkin and Zhou, Yanqi and Jones, Albin and Hashemi, Milad and Swersky, Kevin and Chatterjee, Satrajit and Narayanaswami, Ravi and Laudon, James},
  title     = {{Apollo}: Transferable Architecture Exploration},
  booktitle = {ML for Systems Workshop at NeurIPS},
  year      = {2020},
  url       = {https://research.google/pubs/apollo-transferable-architecture-exploration/}
}

@article{IBMFullSystemSimulation2006,
  author  = {Peterson, James L. and Bohrer, Patrick J. and Chen, Liqun and Elnozahy, Elmootazbellah N. and Gheith, Ahmed and Jewell, Richard H. and Kistler, Michael D. and Maeurer, Theodore R. and Malone, Sean A. and Murrell, David B. and Needel, Neena and Rajamani, Karthick and Rinaldi, Mark A. and Simpson, Richard O. and Sudeep, Kartik and Zhang, Lixin},
  title   = {Application of Full-System Simulation in Exploratory System Design and Development},
  journal = {IBM Journal of Research and Development},
  year    = {2006},
  volume  = {50},
  number  = {2/3},
  pages   = {321--332},
  doi     = {10.1147/rd.502.0321}
}

@inproceedings{Li2026SPAC,
  author    = {Li, Guoyu and Cao, Yang and Ng, Lucas H. L. and Charlton, Alexander and Wang, Qianzhou and Punter, Will and Papaphilippou, Philippos and Guo, Ce and Fan, Hongxiang and Luk, Wayne and Amarasinghe, Saman P. and Brahmakshatriya, Ajay},
  title     = {{SPAC}: Automating {FPGA}-Based Network Switches with Protocol Adaptive Customization},
  booktitle = {2026 IEEE 34th Annual International Symposium on Field-Programmable Custom Computing Machines (FCCM)},
  year      = {2026},
  pages     = {71--80},
  doi       = {10.1109/FCCM68464.2026.00023}
}

@inproceedings{Liu2025SwiftOrExact,
  author    = {Liu, Hang and Geng, Hao and He, Zhuolun and Sun, Qi and Zhuo, Cheng},
  title     = {Swift or Exact? Boosting Efficient Microarchitecture {DSE} via Multi-Fidelity Partial-Order Prediction},
  booktitle = {2025 62nd ACM/IEEE Design Automation Conference (DAC)},
  year      = {2025},
  pages     = {1--7},
  doi       = {10.1109/DAC63849.2025.11133073}
}

@inproceedings{Lo2018MultiFidelityHLS,
  author    = {Lo, Charles and Chow, Paul},
  title     = {Multi-Fidelity Optimization for High-Level Synthesis Directives},
  booktitle = {2018 28th International Conference on Field Programmable Logic and Applications (FPL)},
  year      = {2018},
  pages     = {272--279},
  doi       = {10.1109/FPL.2018.00054}
}

@inproceedings{Monad,
  author    = {Hao, Xiaochen and Ding, Zijian and Yin, Jieming and Wang, Yuan and Liang, Yun},
  title     = {{Monad}: Towards Cost-Effective Specialization for Chiplet-Based Spatial Accelerators},
  booktitle = {2023 IEEE/ACM International Conference on Computer Aided Design (ICCAD)},
  year      = {2023},
  pages     = {1--9},
  doi       = {10.1109/ICCAD57390.2023.10323880}
}

@article{NishiJSSC2023,
  author={Yoshinori Nishi and John W. Poulton and Walker J. Turner and Xi Chen and Sanquan Song and Brian Zimmer and Stephen G. Tell and Nikola Nedovic and John M. Wilson and William J. Dally and C. Thomas Gray},
  title={A 0.297-pJ/Bit 50.4-Gb/s/Wire Inverter-Based Short-Reach Simultaneous Bi-Directional Transceiver for Die-to-Die Interface in 5-nm CMOS},
  journal={IEEE Journal of Solid-State Circuits},
  volume={58},
  number={4},
  pages={1062--1073},
  year={2023},
  doi={10.1109/JSSC.2022.3232024}
}

@misc{OpenAI_Cerebras_2026,
  author       = {{OpenAI}},
  title        = {{OpenAI Partners with Cerebras}},
  year         = {2026},
  month        = jan,
  howpublished = {\url{https://openai.com/index/cerebras-partnership/}},
  note         = {Accessed: Aug. 1, 2026}
}

@article{Polaris,
  author        = {Sakhuja, Chirag and Hong, Charles and Lin, Calvin},
  title         = {{Polaris}: Multi-Fidelity Design Space Exploration of Deep Learning Accelerators},
  year          = {2024},
  journal       = {arXiv preprint arXiv:2412.15548},
  url           = {https://arxiv.org/abs/2412.15548}
}

@ARTICLE{Reticle-limit,
  author={de Diaz, S.L.M. and Fowler, J.W. and Pfund, M.E. and Mackulak, G.T. and Hickie, M.},
  journal={IEEE Transactions on Semiconductor Manufacturing}, 
  title={Evaluating the impacts of reticle requirements in semiconductor wafer fabrication}, 
  year={2005},
  volume={18},
  number={4},
  pages={622-632},
  doi={10.1109/TSM.2005.858502}}

@article{RochaSTCO2024,
  author  = {Giacomini Rocha, Leandro M. and Naeim, Mohamed and Paim, Guilherme and Brunion, Moritz and Venugopal, Priya and Milojevic, Dragomir and Myers, James and Badaroglu, Mustafa and Verhelst, Marian and Ryckaert, Julien and Biswas, Dwaipayan},
  title   = {System--Technology Co-Optimization for Dense Edge Architectures Using {3-D} Integration and Nonvolatile Memory},
  journal = {IEEE Journal on Exploratory Solid-State Computational Devices and Circuits},
  year    = {2024},
  volume  = {10},
  pages   = {125--134},
  doi     = {10.1109/JXCDC.2024.3496118}
}

@inproceedings{Scale-out-processor,
  author    = {Pal, Saptadeep and Petrisko, Daniel and Tomei, Matthew and Iyer, Subramanian S. and Gupta, Puneet and Kumar, Rakesh},
  title     = {Architecting Wafer-Scale Processors---A {GPU} Case Study},
  booktitle = {2019 IEEE International Symposium on High Performance Computer Architecture (HPCA)},
  year      = {2019},
  pages     = {250--263},
  doi       = {10.1109/HPCA.2019.00042}
}

@article{SimulatedAnnealing,
  author  = {Kirkpatrick, Scott and Gelatt, Jr., C. Daniel and Vecchi, Mario P.},
  title   = {Optimization by Simulated Annealing},
  journal = {Science},
  year    = {1983},
  volume  = {220},
  number  = {4598},
  pages   = {671--680},
  doi     = {10.1126/science.220.4598.671}
}

@inproceedings{Sun2021CorrelatedMFOHLS,
  author    = {Sun, Qi and Chen, Tinghuan and Liu, Siting and Miao, Jin and Chen, Jianli and Yu, Hao and Yu, Bei},
  title     = {Correlated Multi-Objective Multi-Fidelity Optimization for {HLS} Directives Design},
  booktitle = {2021 Design, Automation \& Test in Europe Conference \& Exhibition (DATE)},
  year      = {2021},
  pages     = {46--51},
  doi       = {10.23919/DATE51398.2021.9474241}
}

@article{TMAC,
  author  = {Wang, Huizheng and Yang, Qize and Wei, Taiquan and Yu, Xingmao and Li, Chengran and Fang, Jiahao and Lu, Guangyang and Dai, Xu and Liu, Liang and Jiang, Shenfei and Hu, Yang and Yin, Shouyi and Wei, Shaojun},
  title   = {{TMAC}: Training-Targeted Mapping and Architecture Co-Exploration for Wafer-Scale Chips},
  journal = {Integrated Circuits and Systems},
  year    = {2024},
  volume  = {1},
  number  = {4},
  pages   = {178--195},
  doi     = {10.23919/ICS.2024.3515003}
}

@misc{TSMCDTCO2022,
  author       = {Yuan, Lipen},
  title        = {What Is {DTCO}?: An Introduction to Design--Technology Co-Optimization},
  organization = {Taiwan Semiconductor Manufacturing Company},
  howpublished = {{TSMC} Technology Blog},
  year         = {2022},
  month        = jun,
  url          = {https://www.tsmc.com/english/news-events/blog-article-20220615},
  note         = {Accessed: July 31, 2026}
}

@article{Theseus,
  author  = {Zhu, Jingchen and Xue, Chenhao and Chen, Yiqi and Wang, Zhao and Zhang, Chen and Shen, Yu and Chen, Yifan and Cheng, Zekang and Jiang, Yu and Wang, Tianqi and Lin, Yibo and Hu, Wei and Cui, Bin and Wang, Runsheng and Liang, Yun and Sun, Guangyu},
  title   = {{Theseus}: Exploring Efficient Wafer-Scale Chip Design for Large Language Models},
  journal = {IEEE Transactions on Computer-Aided Design of Integrated Circuits and Systems},
  year    = {2025},
  volume  = {44},
  number  = {12},
  pages   = {4793--4806},
  doi     = {10.1109/TCAD.2025.3566297}
}

@inproceedings{Timeloop,
  author    = {Parashar, Angshuman and Raina, Priyanka and Shao, Yakun Sophia and Chen, Yu-Hsin and Ying, Victor A. and Mukkara, Anurag and Venkatesan, Rangharajan and Khailany, Brucek and Keckler, Stephen W. and Emer, Joel S.},
  title     = {{Timeloop}: A Systematic Approach to {DNN} Accelerator Evaluation},
  booktitle = {2019 IEEE International Symposium on Performance Analysis of Systems and Software (ISPASS)},
  year      = {2019},
  pages     = {304--315},
  doi       = {10.1109/ISPASS.2019.00042}
}

@inproceedings{WSC-LLM,
  author    = {Xu, Zheng and Kong, Dehao and Liu, Jiaxin and Li, Jinxi and Hou, Jingxiang and Dai, Xu and Li, Chao and Wei, Shaojun and Hu, Yang and Yin, Shouyi},
  title     = {{WSC-LLM}: Efficient {LLM} Service and Architecture Co-Exploration for Wafer-Scale Chips},
  booktitle = {Proceedings of the 52nd Annual International Symposium on Computer Architecture},
  year      = {2025},
  pages     = {1--17},
  publisher = {ACM},
  doi       = {10.1145/3695053.3731101}
}

@article{Wafer-Scale,
  author  = {Hu, Yang and Lin, Xinhan and Wang, Huizheng and He, Zhen and Yu, Xingmao and Zhang, Jiahao and Yang, Qize and Xu, Zheng and Guan, Sihan and Fang, Jiahao and Shang, Haoran and Tang, Xinru and Dai, Xu and Wei, Shaojun and Yin, Shouyi},
  title   = {Wafer-Scale Computing: Advancements, Challenges, and Future Perspectives},
  journal = {IEEE Circuits and Systems Magazine},
  year    = {2024},
  volume  = {24},
  number  = {1},
  pages   = {52--81},
  doi     = {10.1109/MCAS.2024.3349669}
}

@article{gem5,
  author  = {Binkert, Nathan and Beckmann, Bradford and Black, Gabriel and Reinhardt, Steven K. and Saidi, Ali and Basu, Arkaprava and Hestness, Joel and Hower, Derek R. and Krishna, Tushar and Sardashti, Somayeh and Sen, Rathijit and Sewell, Korey and Shoaib, Muhammad and Vaish, Nilay and Hill, Mark D. and Wood, David A.},
  title   = {{The {gem5} Simulator}},
  journal = {ACM SIGARCH Computer Architecture News},
  year    = {2011},
  volume  = {39},
  number  = {2},
  pages   = {1--7},
  doi     = {10.1145/2024716.2024718}
}

@inproceedings{heterogarnet,
  author    = {Srikant Bharadwaj and Jieming Yin and Bradford Beckmann and Tushar Krishna},
  title     = {Kite: A Family of Heterogeneous Interposer Topologies Enabled via Accurate Interconnect Modeling},
  booktitle = {Proceedings of the 57th ACM/IEEE Design Automation Conference},
  pages     = {1--6},
  year      = {2020},
  doi       = {10.1109/DAC18072.2020.9218539}
}

@inproceedings{ns-3,
  author    = {Khan, Tarannum and Rashidi, Saeed and Sridharan, Srinivas and Shurpali, Pallavi and Akella, Aditya and Krishna, Tushar},
  title     = {Impact of {RoCE} Congestion Control Policies on Distributed Training of {DNNs}},
  booktitle = {2022 IEEE Symposium on High-Performance Interconnects (HOTI)},
  year      = {2022},
  pages     = {39--48},
  doi       = {10.1109/HOTI55740.2022.00021}
}

@inproceedings{WATOS,
  author    = {Wang, Huizheng and Wang, Zichuan and Wang, Hongbin and Hou, Jingxiang and Wei, Taiquan and Li, Chao and Hu, Yang and Yin, Shouyi},
  title     = {{WATOS}: Efficient {LLM} Training Strategies and Architecture Co-Exploration for Wafer-Scale Chip},
  booktitle = {2026 IEEE International Symposium on High Performance Computer Architecture (HPCA)},
  year      = {2026},
  pages     = {1--19},
  doi       = {10.1109/HPCA68181.2026.11408457}
}

@inproceedings{TEMP,
  author    = {Wang, Huizheng and Wei, Taiquan and Wang, Zichuan and Jiang, Dingcheng and Yang, Qize and Liu, Jiaxin and Hou, Jingxiang and Li, Chao and Deng, Jinyi and Hu, Yang and Yin, Shouyi},
  title     = {{TEMP}: A Memory Efficient Physical-Aware Tensor Partition-Mapping Framework on Wafer-Scale Chips},
  booktitle = {2026 IEEE International Symposium on High Performance Computer Architecture (HPCA)},
  year      = {2026},
  pages     = {1--18},
  doi       = {10.1109/HPCA68181.2026.11408568}
}

@inproceedings{ReThermal,
  author    = {Li, Chengran and Wang, Huizheng and Liu, Jiaxin and Liu, Jingyao and Yue, Zhiheng and Li, Xia and Jiang, Shenfei and Deng, Jinyi and Hu, Yang and Yin, Shouyi},
  title     = {{ReThermal}: Co-Design of Thermal-Aware Static and Dynamic Scheduling for {LLM} Training on Liquid-Cooled Wafer-Scale Chips},
  booktitle = {2026 IEEE International Symposium on High Performance Computer Architecture (HPCA)},
  year      = {2026},
  pages     = {1--15},
  doi       = {10.1109/HPCA68181.2026.11408476}
}

@article{PALM,
  author  = {Fang, Jiahao and Wang, Huizheng and Yang, Qize and Kong, Dehao and Dai, Xu and Deng, Jinyi and Hu, Yang and Yin, Shouyi},
  title   = {{PALM}: A Efficient Performance Simulator for Tiled Accelerators with Large-Scale Model Training},
  journal = {arXiv preprint arXiv:2406.03868},
  year    = {2024},
  doi     = {10.48550/arXiv.2406.03868},
  url     = {https://arxiv.org/abs/2406.03868}
}

@inproceedings{SpatialAttention,
  author    = {Wei, Taiquan and Wang, Huizheng and Wang, Zichuan and Yin, Shouyi and Hu, Yang},
  title     = {Spatial-Aware Orchestration of {LLM} Attention on Waferscale Chips},
  booktitle = {Advanced Parallel Processing Technologies: 16th International Symposium, APPT 2025, Athens, Greece, July 13--16, 2025, Proceedings},
  series    = {Lecture Notes in Computer Science},
  volume    = {16062},
  year      = {2026},
  pages     = {386--391},
  publisher = {Springer Nature Singapore},
  doi       = {10.1007/978-981-95-1021-4_29}
}

@article{MOCAP,
  author  = {Wang, Zichuan and Wang, Huizheng and Xiao, Yuheng and Zuo, Haonan and Wei, Taiquan and Deng, Jinyi and Li, Chao and Hu, Yang and Yin, Shouyi},
  title   = {{MOCAP}: Wafer-Scale-Chip-Oriented Memory-Orchestrated Chunked Pipelining Framework for Prefill-Only {LLM} Inference},
  journal = {arXiv preprint arXiv:2606.22968},
  year    = {2026},
  note    = {Accepted to APPT 2026},
  doi     = {10.48550/arXiv.2606.22968},
  url     = {https://arxiv.org/abs/2606.22968}
}

@article{STAR,
  author  = {Wang, Huizheng and Wei, Taiquan and Wang, Hongbin and Wang, Zichuan and Tang, Xinru and Yue, Zhiheng and Wei, Shaojun and Hu, Yang and Yin, Shouyi},
  title   = {Designing Spatial Architectures for Sparse Attention: {STAR} Accelerator via Cross-Stage Tiling},
  journal = {IEEE Transactions on Computers},
  year    = {2026},
  volume  = {75},
  number  = {3},
  pages   = {1125--1140},
  doi     = {10.1109/TC.2025.3648055}
}

@inproceedings{SoWX2025,
  author    = {Shih, Po-Chang and Su, An-Jhih and Tam, King-Ho and Huang, Tze-Chiang and Chuang, Kris and Yeh, John},
  title     = {{SoW-X}: A Novel System-on-Wafer Technology for Next Generation {AI} Server Application},
  booktitle = {2025 IEEE 75th Electronic Components and Technology Conference (ECTC)},
  year      = {2025},
  pages     = {1--6},
  doi       = {10.1109/ECTC51687.2025.00005}
}

@inproceedings{FRED2025,
  author    = {Rashidi, Saeed and Won, William and Srinivasan, Sudarshan and Gupta, Puneet and Krishna, Tushar},
  title     = {{FRED}: A Wafer-Scale Fabric for 3D Parallel {DNN} Training},
  booktitle = {Proceedings of the 52nd Annual International Symposium on Computer Architecture},
  year      = {2025},
  pages     = {34--48},
  publisher = {ACM},
  doi       = {10.1145/3695053.3731055}
}

@inproceedings{James2020PhysicalMapping,
  author    = {James, Michael and Tom, Marvin and Groeneveld, Patrick and Kibardin, Vladimir},
  title     = {{ISPD} 2020 Physical Mapping of Neural Networks on a Wafer-Scale Deep Learning Accelerator},
  booktitle = {Proceedings of the 2020 International Symposium on Physical Design},
  year      = {2020},
  pages     = {145--149},
  publisher = {ACM},
  doi       = {10.1145/3372780.3380846}
}

@inproceedings{Rocki2020Stencil,
  author    = {Rocki, Kamil and Van Essendelft, Dirk and Sharapov, Ilya and Schreiber, Robert and Morrison, Michael and Kibardin, Vladimir and Portnoy, Andrey and Dietiker, Jean-Francois and Syamlal, Madhava and James, Michael},
  title     = {Fast Stencil-Code Computation on a Wafer-Scale Processor},
  booktitle = {SC20: International Conference for High Performance Computing, Networking, Storage and Analysis},
  year      = {2020},
  pages     = {1--14},
  doi       = {10.1109/SC41405.2020.00062}
}

@inproceedings{Yeap2019N5,
  author    = {Yeap, Geoffrey and others},
  title     = {5 nm {CMOS} Production Technology Platform Featuring Full-Fledged {EUV} and High-Mobility-Channel {FinFETs} with Densest 0.021~$\mu$m$^2$ {SRAM} Cells for Mobile {SoC} and High-Performance Computing Applications},
  booktitle = {2019 IEEE International Electron Devices Meeting (IEDM)},
  year      = {2019},
  pages     = {36.7.1--36.7.4},
  doi       = {10.1109/IEDM19573.2019.8993577}
}

@article{Chang2021N5SRAM,
  author  = {Chang, Tsung-Yung Jonathan and Chen, Yen-Huei and Chan, Wei-Min and Cheng, Hank and Wang, Po-Sheng and Lin, Yangsyu and Fujiwara, Hidehiro and Lee, Robin and Liao, Hung-Jen and Wang, Ping-Wei and Yeap, Geoffrey and Li, Quincy},
  title   = {A 5-nm 135-Mb {SRAM} in {EUV} and High-Mobility-Channel {FinFET} Technology with Metal Coupling and Charge-Sharing Write-Assist Circuitry Schemes for High-Density and Low-$V_{\min}$ Applications},
  journal = {IEEE Journal of Solid-State Circuits},
  year    = {2021},
  volume  = {56},
  number  = {1},
  pages   = {179--187},
  doi     = {10.1109/JSSC.2020.3034241}
}

@article{Radhakrishnan2021PowerDelivery,
  author  = {Radhakrishnan, Kaladhar and Swaminathan, Madhavan and Bhattacharyya, Bidyut K.},
  title   = {Power Delivery for High-Performance Microprocessors---Challenges, Solutions, and Future Trends},
  journal = {IEEE Transactions on Components, Packaging and Manufacturing Technology},
  year    = {2021},
  volume  = {11},
  number  = {4},
  pages   = {655--671},
  doi     = {10.1109/TCPMT.2021.3051722}
}

@inproceedings{OrenesVera2024MuchiSim,
  author    = {Orenes-Vera, Marcelo and Tureci, Esin and Martonosi, Margaret and Wentzlaff, David},
  title     = {{MuchiSim}: A Simulation Framework for Design Exploration of Multi-Chip Manycore Systems},
  booktitle = {2024 IEEE International Symposium on Performance Analysis of Systems and Software (ISPASS)},
  year      = {2024},
  pages     = {48--60},
  doi       = {10.1109/ISPASS61541.2024.00015}
}

@article{Duan2024Proteus,
  author  = {Duan, Jiangfei and Li, Xiuhong and Xu, Ping and Zhang, Xingcheng and Yan, Shengen and Liang, Yun and Lin, Dahua},
  title   = {{Proteus}: Simulating the Performance of Distributed {DNN} Training},
  journal = {IEEE Transactions on Parallel and Distributed Systems},
  year    = {2024},
  volume  = {35},
  number  = {10},
  pages   = {1867--1878},
  doi     = {10.1109/TPDS.2024.3443255}
}

@inproceedings{Bang2024vTrain,
  author    = {Bang, Jehyeon and Choi, Yujeong and Kim, Myeongwoo and Kim, Yongdeok and Rhu, Minsoo},
  title     = {{vTrain}: A Simulation Framework for Evaluating Cost-Effective and Compute-Optimal Large Language Model Training},
  booktitle = {2024 57th IEEE/ACM International Symposium on Microarchitecture (MICRO)},
  year      = {2024},
  pages     = {153--167},
  doi       = {10.1109/MICRO61859.2024.00021}
}

@article{Kwon2020MAESTRO,
  author  = {Kwon, Hyoukjun and Chatarasi, Prasanth and Sarkar, Vivek and Krishna, Tushar and Pellauer, Michael and Parashar, Angshuman},
  title   = {{MAESTRO}: A Data-Centric Approach to Understand Reuse, Performance, and Hardware Cost of {DNN} Mappings},
  journal = {IEEE Micro},
  year    = {2020},
  volume  = {40},
  number  = {3},
  pages   = {20--29},
  doi     = {10.1109/MM.2020.2985963}
}

@inproceedings{Jeong2021Union,
  author    = {Jeong, Geonhwa and Kestor, Gokcen and Chatarasi, Prasanth and Parashar, Angshuman and Tsai, Po-An and Rajamanickam, Sivasankaran and Gioiosa, Roberto and Krishna, Tushar},
  title     = {{Union}: A Unified {HW--SW} Co-Design Ecosystem in {MLIR} for Evaluating Tensor Operations on Spatial Accelerators},
  booktitle = {2021 International Conference on Parallel Architectures and Compilation Techniques (PACT)},
  year      = {2021},
  pages     = {30--44},
  doi       = {10.1109/PACT52795.2021.00010}
}

\end{document}